\documentclass[a4paper,11pt]{article}
\pdfoutput=1 % if your are submitting a pdflatex (i.e. if you have
\usepackage{jheppub} % for details on the use of the package, please
\usepackage[all]{xy}

\usepackage[mathlines]{lineno}% Enable numbering of text and display math
\modulolinenumbers[5]% Line numbers with a gap of 5 lines
\title{\boldmath Skyrmion interactions and lattices in chiral magnets: Analytical
    results}

\author[a,b,1]{Calum Ross,\note{Corresponding author.}}
\author[a]{Norisuke Sakai,}
\author[a]{Muneto Nitta,}

\affiliation[a]{Department of Physics and Research and Education Center for Natural Sciences, Keio University,\\ Hiyoshi 4-1-1, Yokohama, Kanagawa 223-8521, Japan}
\affiliation[b]{Department of Physics, University College Cork,\\
Cork, Ireland}

\emailAdd{c.ross@keio.jp}
\emailAdd{norisuke.sakai@gmail.com}
\emailAdd{nitta@phys-h.keio.ac.jp}

\abstract{We study two-body interactions of magnetic skyrmions on the plane 
and apply them to a (mostly) analytic description of a skyrmion lattice. 
This is done in the context of the solvable line, a particular 
choice of a potential for magnetic anisotropy and  Zeeman terms,
where analytic expressions for skyrmions are available. The energy of these analytic single skyrmion solutions is found to become negative below a critical point, where the ferromagnetic state is no longer the lowest energy state. This critical value is determined exactly without the ambiguities of numerical simulations. Along the solvable line the interaction energy for a pair of skyrmions 
is repulsive with power law fall off in contrast to the exponential 
decay of a purely Zeeman potential term. 
Using the interaction energy expressions we construct an inhomogeneous 
skyrmion lattice state, which is a candidate ground states for 
the model in particular parameter regions. Finally we estimate 
the transition between the skyrmion lattice and an inhomogeneous 
spiral state. }

\arxivnumber{2003.07147}

\begin{document} 
\maketitle
\flushbottom

\section{Introduction}
\label{introduction}
Magnetic skyrmions are topologically non-trivial solitons which 
occur in certain magnetic materials \cite{NT}. They were first 
predicted in \cite{BY} as minimum energy configurations of the 
magnetisation vector field and have been observed in a variety 
of magnetic materials \cite{YOKPHMNT,MBJPRNGB,RHMBWVKW}. 
In contrast to the case of baby skyrmions -- field theoretical 
analogues of magnetic skyrmions stabilised by higher derivative 
terms in the energy \cite{PSZ,MS} magnetic skyrmions are stabilised 
by a first order Dzyaloshinskii-Moriya (DM) interaction term 
 (with coefficient $\kappa$) \cite{Dzyaloshinskii,Moriya}. 
The fact that the DM term can give 
a negative contribution to the energy avoids a contradiction 
with the usual Derrick scaling argument \cite{MS}. 
A systematic mathematical study of magnetic skyrmions in the 
presence of an external magnetic field and without anisotropy 
is given in \cite{Melcher}. 
The various phases of a chiral magnet have been explored 
with easy plane anisotropy in addition to an external 
magnetic field \cite{LSB,BRER, APL_thinfilms_2016,MRG_2016}. 
The skyrmion lattice phase has been studied extensively in \cite{BH,RBP}. 
A model of a skyrmion lattice made up of non interacting closely packed 
skyrmions has been considered  in \cite{HZZJN}.
Spiral or helical states have also been studied as effectively 
one-dimensional inhomogeneous states \cite{KO2015,TOPSK_2018}.

Apart from ground breaking experimental studies 
\cite{YOKPHMNT,MBJPRNGB,RHMBWVKW}, almost all theoretical 
studies of magnetic skyrmions are based on numerical simulations 
thus far and more analytic understanding was lacking. 
On the other hand, recently in Ref.~\cite{BRS} a specific potential, 
a combination of a Zeeman term (with coefficient $B$) 
and an easy plane term (with coefficient $A$), was 
considered which results in a solvable model for skyrmions in 
chiral magnetic thin films.
More explicitly, tuning the two coefficients of the potential 
to be $B=2A$, one obtains exact hedgehog skyrmion solutions with 
topological charge (degree as a mapping $S^{2}\to S^{2}$) $Q=-1$. 
This relation $B=2A$ of parameters is called the solvable line. 
These exact skyrmion solutions are a natural generalisation 
of the ferromagnetic skyrmions in the Heisenberg model \cite{BP}. The holomorphic properties of the skyrmion solutions along the solvable line were first discussed in \cite{MD2016}. While the anisotropy, $A$, and the DM constant, $\kappa$, are determined by the material the external magnetic field, $B$, can be tuned. This suggests that the solvable line should be accessible in real materials through a careful tuning of the external magnetic field. 

Along the solvable line $B=2A$, there is a special point tuned 
to the coefficient $\kappa$ of the DM interaction through $B=2A=\kappa^2$. 
This is a so-called Bogomol'nyi-Prasad-Sommerfield 
(BPS) \cite{Bogomolny:1975de,Prasad:1975kr} point where the 
model has an infinite number of analytic solutions constructed 
in terms of holomorphic functions. 
Such BPS limits were first discussed in the context of monopoles 
in quantum field theories, and are now known to appear in 
various theories admitting topological solitons \cite{MS}, in particular in supersymmetric 
field theories (see, e.~g., \cite{Eto:2006pg,Shifman:2009zz}). 
In the case of superconductors, it corresponds to the critical coupling 
between type-I and type-II superconductivities. Some further studies of BPS magnetic skyrmions and instantons in chiral magnets
have been made in Refs.~\cite{Adam:2019yst, Adam:2019hef, Hongo:2019nfr}.

The exact skyrmion solution along the solvable line reveals the subtlety that a boundary contribution to the energy functional is required for 
the DM interaction term to be well defined \cite{BRS,Schroers4,Schroers5}. 
This boundary contribution is closely related to the slow power law asymptotic decay 
of the magnetisation vector for the skyrmion solution. 
This boundary contribution can give a negative contribution to the energy of the exact skyrmion solution, which can result in the skyrmion having lower energy than
the homogeneous ferromagnetic state. 
This fact opens up the possibility of understanding the 
inhomogeneous ground state, such as skyrmion lattice states,
if a balance is met between the negative energy of individual 
skyrmions and the interaction energy between them. It is worth emphasising that exact single skyrmion solutions are only known for the solvable line and not for the better studied region $B>2A$. Outside the solvable line, such as for $B>2A$, numerical analysis is needed to construct skyrmion solutions.

The asymptotic decay of the skyrmion fields is also a key difference with a power law decay for the solvable line and an exponential decay for $B>2A$. The slower, power law, decay for skyrmions along the solvable line mean that the asymptotic methods used for $B>2A$ are not as easily applicable. This is because boundary terms in the energy, which vanish for exponential decay, play an important role along the solvable line. Even though $B=2A$ has the subtlety of boundary terms, the presence of exact solutions overrides this disadvantage.

The interaction energy between a pair of skyrmions was discussed for a specific easy-axis potential in \cite{Bogdanov1995} and has recently been discussed \cite{LRSB_2013,CGC,BLH}, using the dipole 
approximation introduced in Ref.~\cite{PSZ}. 
Most previous work, including Refs.~\cite{CGC,BLH}, only considers the case of an external magnetic field without an anisotropy term. It has been found that the 
interaction energy decays exponentially at large separations 
in this case \cite{FKATDS}. 
The case of easy plane anisotropy being more important than 
the Zeeman term was considered in Ref.~\cite{LK2017} 
where an attractive interaction was found between individual skyrmions.  The skyrmion interaction energy for the solvable line does not appear to have been studied before. The boundary contributions to the energy caused by the slow, power law, decay of the skyrmion configurations complicate the asymptotic approach used for $B>2A$. This necessitates a slightly different approach which is possible due to the presence of analytic skyrmion solutions.

In this paper we conduct a semi-analytic study of interactions 
between magnetic skyrmions along the solvable line, and apply 
it to construct a model of a skyrmion lattice. 
To achieve this, we make use of the dipole approximation from 
\cite{PSZ}. This approximation involves working with a superposition 
of two skyrmions, and is reliable as long as the skyrmions are well separated, 
becoming less reliable as the distinction between the constituent 
skyrmions becomes less pronounced. 
The accuracy of the approximation should be measured in 
terms of the separation between two skyrmions divided by the 
diameter of a single skyrmion. 
The diameter of the skyrmion is determined by three dimensionful 
parameters, the DM interaction strength $\kappa$, and the two parameters 
$A$ (for magnetic anisotropy) and $B$ (for a Zeeman field) in 
the potential. 
 From the numerical study of the phase diagram of a chiral magnet with 
easy plane anisotropy, \cite{LSB,BRER}, it is known that large 
values of $A$ and $B$ correspond to homogeneous ferromagnetic phases. 
The solvable line, $2A=B$, lies along the phase boundary between 
the polarised ferromagnetic phase and the spin canted ferromagnetic phase. 
At large values of the potential along the solvable line, magnetic 
skyrmions are obtained as positive energy excited states on top of the 
ferromagnetic ground state. The skyrmions have smaller radii as 
the potential parameter $A$ becomes larger. 
Thus the dipole approximation, at a fixed separation between 
two isolated skyrmions, works best as the potential parameter $A$ 
becomes large. 
For small values of $A$, on the other hand, the skyrmions along 
the solvable line have negative energy and the homogeneous 
ferromagnetic state becomes unstable. This leads to the possibility 
of the ground state being an inhomogeneous skyrmion lattice in 
agreement with numerical studies \cite{BH,RBP}. The other possible 
inhomogeneous state is a spiral or helical state 
\cite{KO2015,TOPSK_2018} where 
the configuration is effectively one dimensional.

We apply the dipole approximation at smaller values of $A$ to 
understand some of the qualitative behaviour of the triangular 
skyrmion lattice phase from \cite{LSB} along the solvable line. 
We take two exact single skyrmion solutions and consider 
their superposition as a function of their separation. 
Our results show that the dipole 
approximation can be used reliably to find the transition 
between the ferromagnetic phase and the lattice phase. 
However, the dipole approximation breaks down before 
we can study the phase transition between the lattice phase 
and the spiral phase. We find evidence that a model of a 
skyrmion lattice as, effectively, non interacting closely packed skyrmions \cite{HZZJN} 
becomes a better approximation.  
Using this we estimate the critical value of the external 
magnetic field where the transition to the spiral phase occurs.

The key feature of the present work is that by working on 
the solvable line there are exact expressions for the skyrmion 
configurations enabling us to make exact statements about the energy of a single skyrmion. This makes it possible for us to identify a specific value of the anisotropy below which the single skyrmion solution has negative energy and we have a state lower in energy than the ferromagnetic state. Combining the exact single skyrmion configurations with the dipole approximation we arrive at an analytic expression for the interaction 
energy density of the superposition without the need to work 
asymptotically. After numerical integration we find that the interaction energy is positive 
and has a power law decay in contrast to the exponential decay 
of the pure Zeeman potential \cite{FKATDS,CGC,BLH}. An exponentially decaying interaction was also found in \cite{Bogdanov1995} for the case of a specific easy-axis potential. Comparing these previous studies with our Appendix  \ref{boundary term discussion}, we see that the interaction energy will exponentially decay when the skyrmion field decays exponentially. Two-body and three-body skyrmion interactions in the absence of an anisotropy term in the potential have been studied in \cite{TSA2019}.
We show that the power law decay of the skyrmion fields necessitates the inclusion of a boundary 
term in the energy functional in order to have a well defined variational problem when deriving the equations of motion. Along the solvable line, we find a repulsive interaction.

This paper is organized as follows.
In Sec.~\ref{sec:model}, we review solvable models for 
magnetic skyrmions and clarify the necessity of a boundary term.
In Sec.~\ref{sec:interaction}, we calculate the interaction energy 
and force between two isolated skyrmions.
In Sec.~\ref{sec:lattice}, the interaction energy is applied to 
describe a skyrmion lattice, and the phase transition to the spiral 
phase is estimated.
Sec.~\ref{sec:summary} is devoted to a summary and discussion.
Appendix \ref{boundary term discussion} gives the details of 
the boundary term in the energy functional and its relation to the
asymptotic behaviour of the magnetisation vector field.
In Appendix \ref{interaction energy density computation} we express the 
interaction energy as a function of dimensionless variables. 
Spiral states are described in Appendix \ref{spiral state appendix}.

\section{Solvable models for magnetic skyrmions}\label{sec:model}

In this section, we review solvable models for magnetic skyrmions.
In Subsec.~\ref{sec:solvmodel}, we introduce the model and classify 
homogeneous ground states, when the DM term can be neglected.
Then in Subsec.~\ref{sec:exact}, we present exact skyrmion solutions. 
Finally in Subsec.~\ref{sec:instability}, we discuss the instability of 
the ferromagnetic phase.

\subsection{The model}\label{sec:solvmodel}

This subsection gives a concise summary of previously 
known results that are directly relevant for our study and 
clarifies our conventions at the same time. 
We study chiral magnets in two spatial dimensions 
described in terms of a real three component magnetisation 
vector field $\vec{n}:\mathbb{R}^{2}\to S^{2}$, constrained 
as $\vec{n}^2=1$. 
In works such as \cite{BY,BH,LSB} the following energy functional 
is considered 
%\begin{widetext}
\begin{equation}
E[\vec{n}]_{\rm bulk}
=\int_{\mathbb{R}^{2}} \left[\frac{1}{2}|\nabla \vec{n}|^{2}
+\kappa\;%\left(
\vec{n}%-e_{3}\right)
\cdot \left(\nabla_{-\alpha}\times 
\vec{n}\right)+U(n_{3})\right]d^{2}x, 
\label{energy-functional-general}
\end{equation}
where the potential $U$ consists of a Zeeman term with the magnetic 
field $B$ as coefficient, and an anisotropy term with the 
coefficient $A$ 
\begin{equation}
U(n_{3})=-B n_{3}+An_{3}^{2}. 
\label{Lin potential}
\end{equation}
The term corresponding to the parameter $\kappa$ is
the Dzyaloshinskii-Moriya (DM) interaction energy 
\cite{Dzyaloshinskii,Moriya}, and the rotated derivative is defined as 
\begin{equation}
\nabla_{-\alpha}=
\left(
\begin{array}{c}
\cos\alpha \partial_1+\sin\alpha \partial_2 \\
-\sin\alpha \partial_1+\cos\alpha \partial_2 \\
0
\end{array}
\right) ,
\label{eq:DMderivative}
\end{equation}
where the parameter $\alpha\in [0,2\pi)$ denotes a one parameter 
family of models corresponding to different types of DM term, 
for example $\alpha=0$ corresponds to the Bloch type DM and 
$\alpha=\frac{\pi}{2}$ corresponds to the N\'eel type DM interaction.
The subscript ``bulk'' indicates that we need to add a boundary 
term to properly define the DM interaction term (to apply the variational 
principle), as we describe subsequently. 

\begin{figure}[htbp]
\begin{center}
\includegraphics[width=0.45\textwidth]{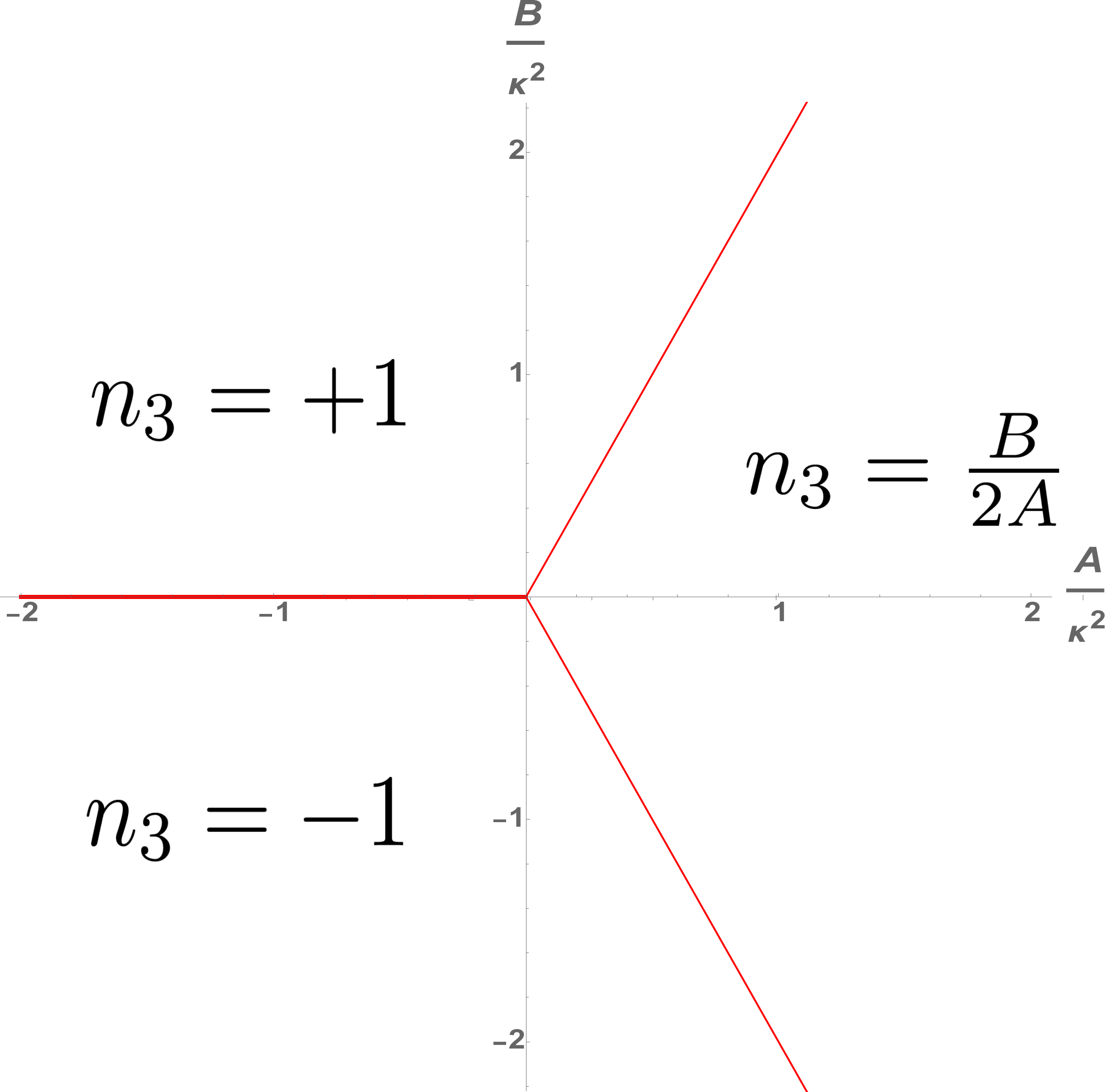}
\caption{The three ferromagnetic phases of the potential $U(n_{3})
=-B n_{3}+An_{3}^{2}$.}
\label{phase fig}
\end{center}
\end{figure}

For large values of the potential parameters $|A|, |B| \gg \kappa$, 
for which the DM term can be neglected, the ground state is 
determined by the minimum of the potential. 
There are three types of minima of the general potential $U(n_3)$ 
depending on values of $A, B$:
\
\begin{enumerate}
\item
$\vert B\vert<2A$: canted ferromagnetic phase ($n_3=\frac{B}{2A}$), 

\item
$B>2A,\, B>0$: Ferromagnetic phase ($n_3=+1$),

\item
$B<-2A,\, B<0$: Ferromagnetic phase ($n_3=-1$).

\end{enumerate}
These three phases exhaust the possible homogeneous ground states. 
The phase diagram for the ground states of the potential is 
included in Figure \ref{phase fig}.

Let us note that the model is invariant under the transformation 
\begin{equation}
B \to -B, \quad n_3 \to -n_3 .
\label{eq:inversion_B}
\end{equation}
We consider the parameter region $B\ge 0$ in the following, since 
we can obtain identical result for $B<0$ by just changing 
the sign of $n_3$.

If we tune the parameters $A, B$ in the potential as 
\begin{equation}
B=2A, 
\label{eq:solvable}
\end{equation}
(and add a constant $A$), we obtain a potential $V(n_3)$ instead of 
$U(n_3)$ (in the case of $B<0$, we should tune $-B=2A$ to obtain a 
potential $V(n_3)=A\left(1+n_{3}\right)^{2}$ instead) 
\begin{equation}
V(n_3)=A\left(1-n_{3}\right)^{2} .
\label{eq:potential_solvable}
\end{equation}
We call this choice of parameters, $B=2A$, the solvable line, since explicit skyrmion configurations can be constructed 
as shown in the next subsection.

Instead of the constrained vector field $\vec{n}$ with $\vec{n}^2=1$, 
it is often more convenient to use the unconstrained complex field $w$ 
defined by the stereographic projection 
\begin{equation}
w=\frac{n_{1}+in_{2}}{1+n_{3}}. 
\label{complex field}
\end{equation}
The energy functional in Eq.~\eqref{energy-functional-general} 
with the potential replaced by $V$ in Eq.~\eqref{eq:potential_solvable} 
is rewritten as 

\begin{equation}
E[w]_{\rm bulk}^{B=2A}
=\int_{\mathbb{R}^{2}} \frac{2|\nabla w|^2
+4\kappa {\rm Im}(e^{i\alpha}(\partial_z w+ w^2\partial_z\bar{w})) 
+4A|w|^4}{(1+|w|^2)^2}d^{2}x . 
\label{bulk-energy-functional-w}
\end{equation}

%%%%%%%%%%
\subsection{Exact single skyrmion solutions}\label{sec:exact}

Defining isolated skyrmion states requires knowledge of the 
minima of the potential. This is because the magnetisation vector 
of the skyrmion configuration is required to go to the ground 
state asymptotically. 

In the case of derivative interactions like the DM term, 
particular care is needed to properly define the energy 
functional. When varying fields to obtain field equations, 
we need to integrate by parts, which produces a surface term. 
If the fields do not decay sufficiently 
fast asymptotically, the surface term obstructs the variational 
principle used to obtain the field 
equation. 
Hence it is mandatory to subtract a boundary term to remove 
this obstruction and allow us to apply the variational principle. 
In the case of the DM interaction,  
this is accomplished by adding the following boundary term 
\cite{Schroers4,Schroers5} to 
the energy functional \eqref{energy-functional-general}
\begin{equation}
-4\pi\Omega[\vec{n}]=-\kappa \int_{\mathbb{R}^{2}}
\vec{n}_{\infty}\cdot \left(\nabla_{-\alpha}\times \vec{n}\right)d^{2}x. 
\label{total vortex strength}
\end{equation}  
Here $\vec{n}_{\infty}$ is the asymptotic value of $\vec{n}$ 
determined by the homogeneous state. Explicitly, for the ferromagnetic 
phases, $\vec{n}_{\infty}=\pm e_{3}$ with the top sign when $B$ 
is positive and the lower sign when $B$ is negative. 

We wish to stress that only by adding the boundary term,
does the DM term become well-defined in terms of the variational 
problem for the energy functional.  
More details about the necessity of the boundary term are discussed 
in Appendix \ref{boundary term discussion}. 
Some properties of the boundary term in Eq.~\eqref{total vortex strength}, 
are discussed in Refs.~\cite{BRS,Melcher}.

Taking into account the boundary term in 
Eq.~\eqref{total vortex strength} and the potential in 
Eq.~\eqref{eq:potential_solvable}, the model of a chiral magnet 
that we consider is 
\begin{equation}
E[\vec{n}]=E_{\rm bulk}^{B=2A}-4\pi\Omega =\int_{\mathbb{R}^{2}} 
\left[\frac{1}{2}|\nabla \vec{n}|^{2}
+\kappa\;\left(
\vec{n}-e_{3}\right)
\cdot \left(\nabla_{-\alpha}\times 
\vec{n}\right)+V(n_3)%\frac{B}{2}\left(1-n_{3}\right)^{2}
\right]d^{2}x. 
\label{energy functional static}
\end{equation}
The tuning of the potential allows one to obtain exact solutions 
of a hedgehog type \cite{BRS}. In terms of the polar coordinates 
$x_1=r\cos\varphi$, $x_2=r\sin\varphi$, one finds the exact 
solution by imposing a boundary condition $\vec{n}\to (0,0,1)$ 
at infinity $r\to\infty$,
\begin{equation}
\vec{n}=\begin{pmatrix}
\sin\Theta \cos\left(\varphi +\gamma\right)\\
\sin\Theta \sin\left(\varphi +\gamma\right)\\
\cos\Theta
\end{pmatrix}\label{Hedgehog solution}
\end{equation}
with phase $\gamma=\frac{\pi}{2}-\alpha$ and profile function 
\begin{equation}
\Theta=2\arctan\left(\frac{\kappa}{A r}\right). 
\label{profile function}
\end{equation}

The hedgehog solution is called a skyrmion. 
It gives a mapping from the base space ${\mathbb{R}^{2}}$ 
to the target space $S^{2}$. As $\vec{n}$ tends to a constant as 
$r\to \infty$ we can regard ${\mathbb{R}^{2}}$ 
as $S^2$ by adding the point at infinity, thus we can define the degree $Q$ 
corresponding to $\pi_2(S^2)$  
\begin{align}
Q[\vec{n}]
=&\frac{1}{4\pi}\int \vec{n} \cdot \partial_1\vec{n} 
\times \partial_2\vec{n}\; d^2 x .
\label{eq:skyrmion_charge}
\end{align}
This topological charge is $Q=-1$ for a skyrmion
configuration.
It is worth repeating that in this context a skyrmion 
has topological charge $Q=-1$ while an anti-skyrmion has topological 
charge $Q=+1$. This is in contrast to the case of baby skyrmions
as discussed in \cite{MS}.

The skyrmion solution has energy
\begin{equation}
E=4\pi\left(1-\frac{\kappa^{2}}{A}\right).
\label{hedgehog energy expression}
\end{equation}
Here, when calculating the energy of the skyrmion, taking 
into account the boundary term (\ref{total vortex strength}) 
is crucial. This is an exact expression for the skyrmion energy 
as a function of $\frac{A}{\kappa^{2}}$ along the solvable line. 
The analytic skyrmion solutions, Eqs.~\eqref{Hedgehog solution} 
and~\eqref{profile function}, along the solvable line enable 
us to compute this exact energy expression.
As highlighted in Table \ref{energy table}, subtracting the 
boundary term leads to the skyrmion having negative energy 
which is very important for our approach to constructing the 
skyrmion lattice.
%%%%%%%%
\begin{table}
\begin{center}
\caption{\label{energy table} A table comparing the energy of 
the exact skyrmion solution and BPS anti-skyrmion with and without 
the boundary term, Eq.~\eqref{total vortex strength}. The BPS 
case is $B=2A=\kappa^{2}$ and the anti-skyrmion is constructed 
from Eq.~\eqref{bps skyrmions} with $f(z)=z$. Subtracting the 
boundary term is necessary for the skyrmions to have negative energy.}
\begin{tabular}{|c|c|c|c|}
\hline
& Skyrmion &BPS skyrmion& BPS anti-skyrmion  \\ \hline
$E_{\rm bulk}^{B=2A}$&$4\pi $& $4\pi$& $4\pi$ \\ \hline
$E[\vec{n}]$&$4\pi\left(1-\frac{\kappa^{2}}{A}\right)$&$-4\pi$ & $4\pi$ 
\\ \hline
\end{tabular}
\end{center}
\end{table} 
%%%%%%%%%
The profile function in Eq.~\eqref{profile function} shows that 
$n_{3}=0$ (the magnetisation vector lies in the $x-y$ plane) 
at 
\begin{equation}
r=\frac{\kappa}{A}, 
\label{eq:radius}
\end{equation}
which is defined as the radius of a skyrmion.

In terms of the complex field $w$ of the stereographic projection, 
the energy functional and topological charge can be expressed as 
\begin{eqnarray}
E[w]=4\pi Q+
\int_{\mathbb{R}^{2}} \frac{8|\partial_z \bar w|^2
+8\kappa {\rm Im}(e^{-i\alpha}w^2\partial_z\bar w)+4A|w|^4}
{(1+|w|^2)^2}d^{2}x, 
\label{energy functional-w}
\end{eqnarray}
and
\begin{equation}
Q=\frac{1}{\pi}\int_{\mathbb{R}^{2}} 
\frac{\partial_z w\partial_{\bar z}\bar w 
-\partial_{\bar z} w\partial_{z}\bar w} 
{(1+|w|^2)^2}d^{2}x,
\label{degree-w}
\end{equation}
respectively.
The single skyrmion exact solution can be 
rewritten in terms of the stereographic complex field 
$w$ in Eq.~\eqref{complex field} as  
\begin{equation}
w=i\frac{\kappa}{A}\frac{e^{-i\alpha}}{\bar{z}-\bar{b}}  ,
\label{hedghog sol in complex coords}
\end{equation}
where the centre of the skyrmion is denoted as $z=b$, here
$\vec{n}=(0,0,-1)^T$.

In Ref.~\cite{BRS} it was shown that for the BPS case ($B=2A=\kappa^2$) 
we can complete the square in the energy functional 
\eqref{energy functional static} and find first order BPS 
equations 
\begin{equation}
D_{1}\vec{n}=-\vec{n}\times D_{2}\vec{n}, 
\qquad D_{2}\vec{n}=\vec{n}\times D_{1}\vec{n}
\end{equation}
where $D_{i}\vec{n}=\partial_{i}\vec{n}-\kappa e^{-\alpha}_{i}\times\vec{n}$, 
for $i=1,2$, is a covariant (helical) derivative. 
In terms of the stereographic field $w$ in Eq.~\eqref{complex field},
the BPS equations become 
\begin{equation}
\partial_{\bar{z}}w=\frac{i}{2}e^{i\alpha}\kappa w^{2}, \label{complex BPS}
\end{equation}
and there are infinitely many BPS solutions, with topological 
charge $Q\geq -1$, of the form
\begin{equation}
w=\frac{2i}{e^{i\alpha}\kappa\bar{z}+2if(z)}, \label{bps skyrmions}
\end{equation}
for $f(z)$ an arbitrary holomorphic function. 
Solutions of these BPS  
equations with the boundary 
condition $n_3\to 1$ at $r\to\infty$ have energy given by 
their topological charge 
$E[\vec{n}]=4\pi Q[\vec{n}]=\int \vec{n}\cdot 
\left(\partial_{1}\vec{n}\times \partial_{2}\vec{n}\right)d^{2}x$. 
This is in contrast to the usual story for soliton solutions 
in the familiar $O(3)$ sigma model \cite{MS} where the energy 
is given by $4\pi \vert Q[\vec{n}]\vert$.
The exact Hedgehog solution in Eq.~(\ref{hedghog sol in complex coords}) 
at $2A=\kappa^2$ coincides with the BPS solution with $Q=-1$. 
Full details of these BPS solutions are given in Refs.~\cite{BRS,Schroers4}.

\subsection{Instability of ferromagnetic phase 
}\label{sec:instability}

We are now in a position to consider the ground state of the 
chiral magnet described by the energy functional in 
\eqref{energy functional static} along the solvable line 
$B=2A> 0$. 
The solvable line coincides precisely with the boundary between the 
ferromagnetic ($n_3=+1$) and canted ferromagnetic phases. 
Along the solvable line, we obtain the ferromagnetic 
ground state $n_3=+1$ in the limit $A\to \infty$. 
If we decrease the potential strength parameter $A$, the DM 
interaction becomes more important and can result in a negative 
energy. 
The exact single skyrmion energy Eq.~\eqref{hedgehog energy expression} shows that a single skyrmion 
is a positive energy excited state above the ferromagnetic 
ground state for $A>\kappa^2$. 
On the other hand, the negative energy of the exact single 
skyrmion solution for $A<\kappa^2$ reveals that the homogeneous 
ferromagnetic state is no longer the lowest energy state.  
As $A$ decreases down to $\kappa^{2}$, we expect that skyrmions will start to 
appear as defects in the homogeneous ferromagnetic state. 
Therefore we expect that there is a phase boundary 
at $A=\kappa^2$, below which an inhomogeneous state 
with many skyrmions is favored over the homogeneous ferromagnetic 
state. Repeating this for emphasis, the analytic skyrmion solutions become negative energy states for $A<\kappa^{2}$ and thus the ferromagnetic state is no longer the lowest energy solution of the equations of motion. This suggests that there will be a phase transition to a new inhomogeneous ground state constructed from the negative energy skyrmions. This is our first observation of the precise value of 
the phase transition point in a chiral magnet. This transition point cannot be established precisely by numerical methods and shows the power of the analytical single skyrmion solutions.

Detailed numerical studies \cite{LSB,APL_thinfilms_2016,MRG_2016} showed that 
the inhomogeneous phases are the skyrmion lattice phase, at 
intermediate values of $A$ below $\kappa^2$, 
and the spiral phase, for the region much below $\kappa^2$. 
In subsequent sections, we wish to shed more light on the 
nature of these inhomogeneous ground 
states and their phase boundaries by analytic means, rather than numerically.

As we have an exact expression for the energy 
\eqref{hedgehog energy expression} and the radius 
\eqref{eq:radius} of a skyrmion, we can  compute the  
energy density inside a  skyrmion. 
We assume that the skyrmion energy \eqref{hedgehog energy expression} 
is approximately contained inside the circle of the skyrmion 
radius $r=\kappa/A$ with the area 
$\pi \left({\kappa}/{A}\right)^{2}$. 
This leads to the skyrmion energy density being
\begin{equation}
\langle e_{\text{Sk}}\rangle=\frac{4\pi
\left(1-\frac{\kappa^{2}}{A}\right)}
{\pi \left(\frac{\kappa}{A}\right)^{2}}
=4 A\left( \frac{A}{\kappa^{2}} -1 \right).
\label{eq:average_energy_density}
\end{equation}
We observe that the energy density is symmetric around 
$A=\frac{\kappa^{2}}{2}$.

For values of $A$ higher than $\kappa^{2}$, the energy density 
is positive. 
As $A\to \infty$, the single skyrmion energy increases towards 
$4\pi$ and the skyrmion radius goes to zero. 
Conversely, the skyrmion radius diverges in the limit $A\to 0$, 
indicating that the $n_{3}=0$ circle becomes very large and that 
the region where $n_{3}=-1$  covers most of the interior.  
In the same limit the energy of a single skyrmion diverges. 
However, the average energy density \eqref{eq:average_energy_density} 
goes to zero.

\section{Interactions of magnetic skyrmions}\label{sec:interaction}

Although it is energetically more favorable to produce skyrmions 
below the phase transition point, one cannot create infinitely 
many skyrmions because of their interaction. Therefore, 
it is important to study the interactions between skyrmions in 
order to understand the inhomogeneous ground state. 
We can use the method from Refs.~\cite{PSZ,FKATDS} and 
consider the case of two $Q=-1$ skyrmions. 
The centres of the skyrmions are defined to be the points where 
$n=\begin{pmatrix}0,&0,&-1\end{pmatrix}^{T}$. 
In this section we explain the approximation and use it to 
fit the interaction energy as a function of the ratio of separation 
to the skyrmion diameter for various values of $\frac{A}{\kappa^2}$.

\subsection{Superpositions of skyrmions}
We study the interaction between two magnetic skyrmions of degree 
$Q=-1$ by constructing a degree $Q=-2$ skyrmion configuration, 
which approaches a solution of the field equations at infinite 
separations. In Refs.~\cite{PSZ,FKATDS}, the magnetisation field 
$\vec{n}$ is used to superpose two skyrmion solutions, $\vec{n}^{A}$ 
and $\vec{n}^{B}$. The degree of the superposition is the sum 
of the degrees of $\vec{n}^{A}$ and $\vec{n}^{B}$, in this case 
$Q=-2$. Here, we work at the level of the complex field in 
Eq.~\eqref{complex field} where the superposition is 
\begin{equation}
w=w_{A}+w_{B} .\label{complex superposition}
\end{equation}

\begin{figure}[htbp]
\begin{center}
\includegraphics[width=0.3\textwidth]{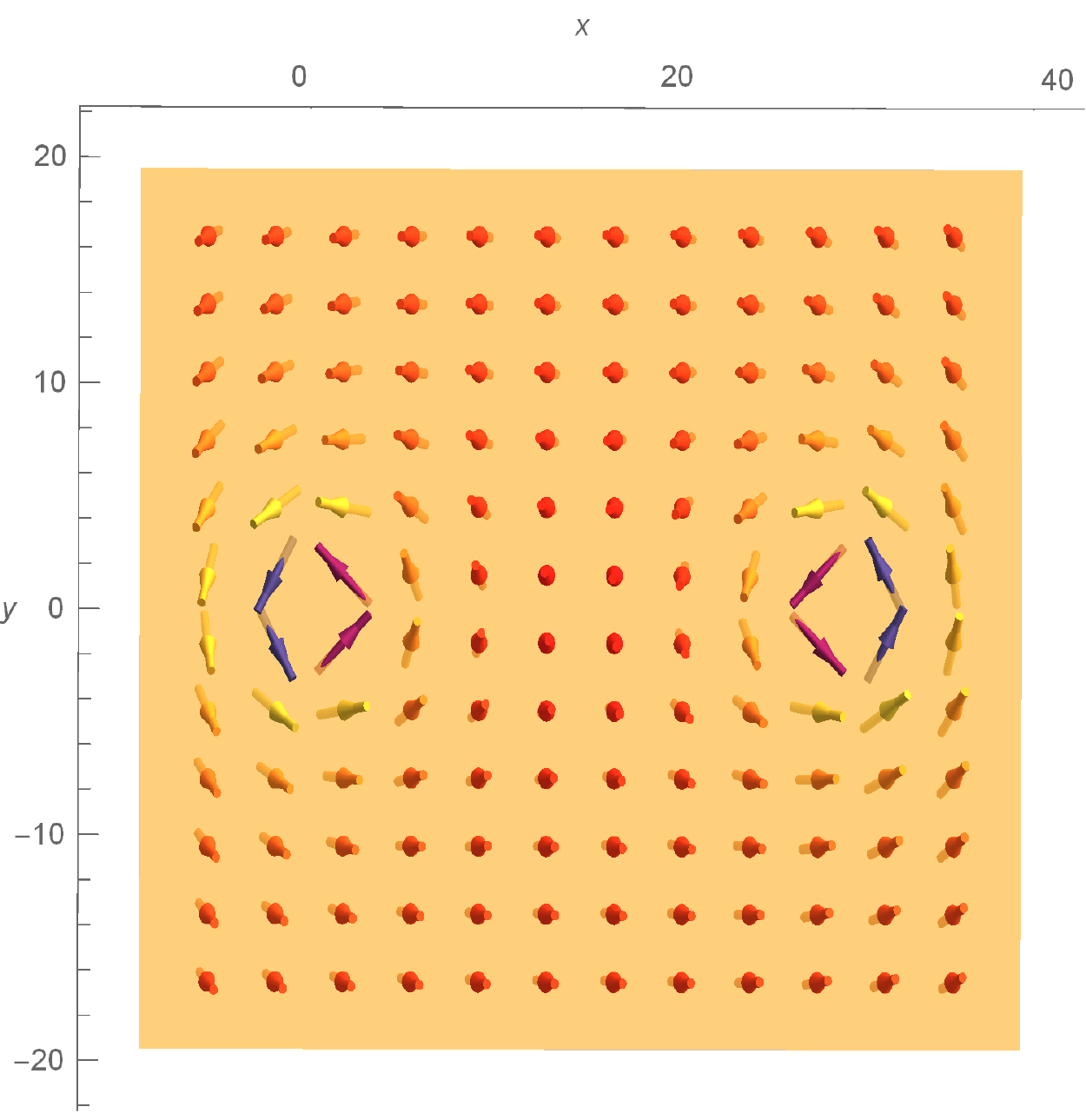}
\includegraphics[width=0.3\textwidth]{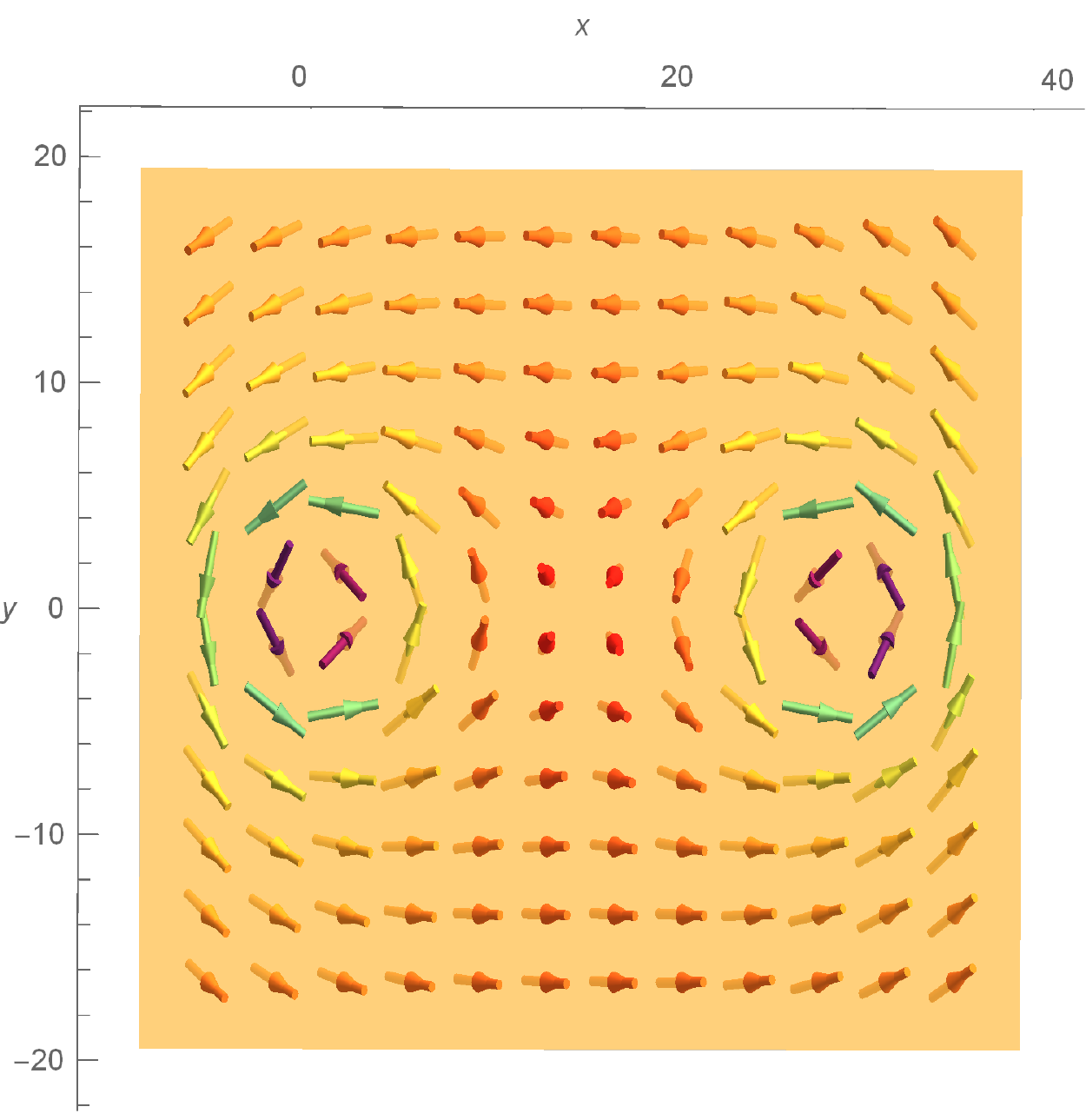}
\includegraphics[width=0.3\textwidth]{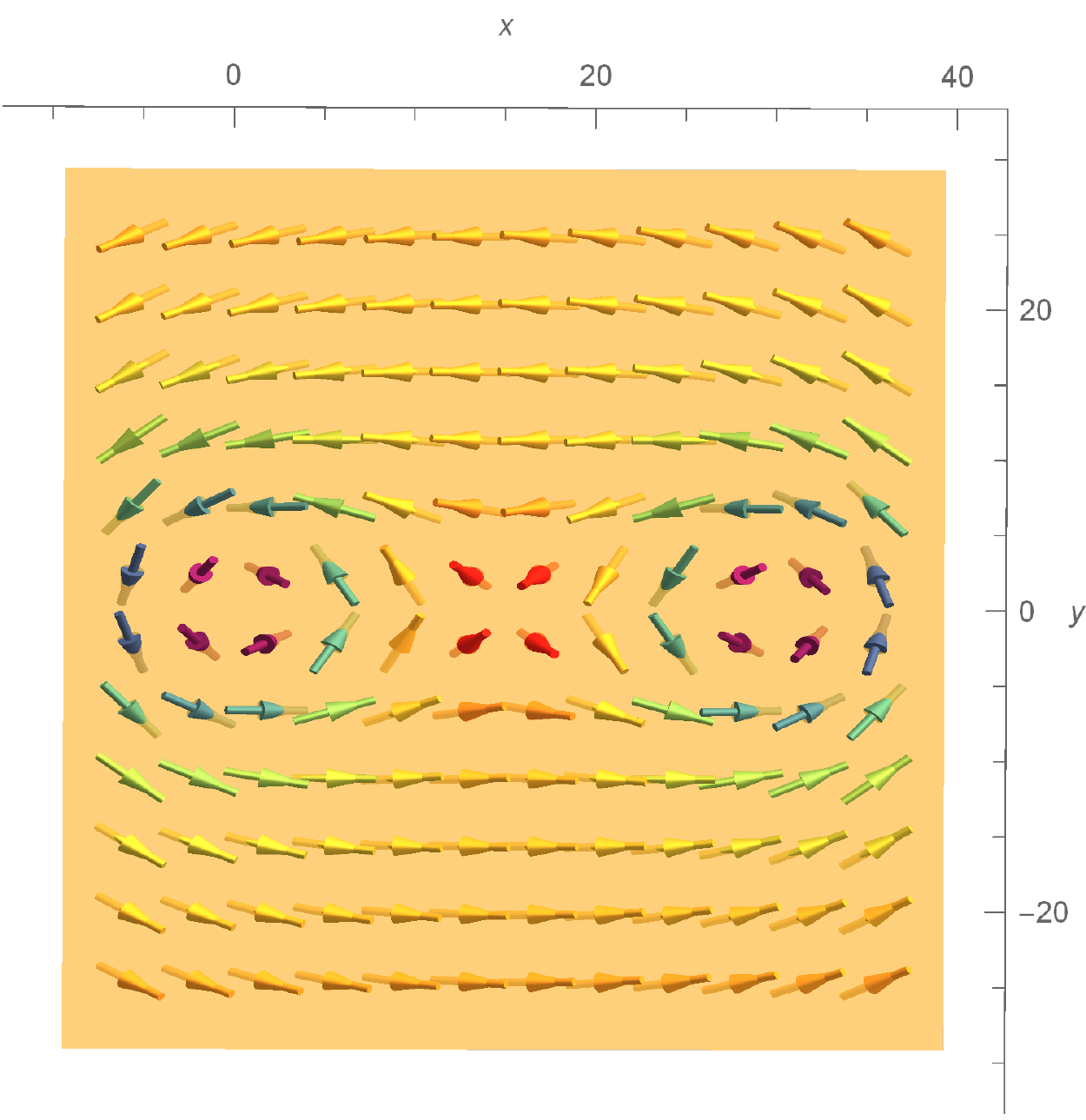}
\caption{The magnetisation vector field for the superposition of two $Q=-1$ hedgehog skyrmions  with one skyrmion centred at zero and the other skyrmion centred at $(30\kappa^{-1},0)$. On the left the plot is for $A=\frac{\kappa^{2}}{2}$, in the middle is $A=\frac{2}{10}\kappa^{2}$, and on the right is $A=\frac{1}{10}\kappa^{2}$. This shows that as $A$ decreases, at fixed separation, the interpretation of the superposition as two distinct skyrmions disappears and the approximation becomes worse.}
\label{superposition magnetisation vector fig}
\end{center}
\end{figure}

The superposition in terms of the complex field has the advantage 
of automatically satisfying the constraint $\vec{n}^{2}=1$, in 
contrast to the superposition in terms of $\vec{n}$.
The two degree $Q=-1$ fields, $w_{A}$ and $w_{B}$, both solve 
the equations of motion. However, the superposition $w$ does 
not as the equations of motion are non-linear. 
As stated above the superposition becomes a solution only in the 
limit of infinite separation. 
Hence the accuracy of the approximation improves as the centres 
of the skyrmions (the poles of the $w$'s) get further apart. 
Some examples of these superposition configurations at different 
values of $A$ are included in 
Fig.~\ref{superposition magnetisation vector fig}. 
These figures show that as $A$ decreases the approximation becomes 
less reliable. In fact the key quantity to consider is the ratio 
of the separation to the skyrmion diameter, this will be made 
explicit below. For the case $B>2A$ where the skyrmion fields 
are exponentially decaying, the decay exponent gives a characteristic 
length scale which controls the validity of the dipole approximation. 
Along the solvable line the power law decay complicates matters 
and it is not as obvious how to extract a characteristic length 
scale. However, it is still known that the superposition approaches 
a true solution for infinite separation. This means that the 
accuracy of the approximation will improve with increasing separation. 
In fact we show below that it is best to work in terms of a 
dimensionless separation which demonstrates that the accuracy 
of the approximation also depends on the value of $A$.

We take a skyrmion $w_A$ centred at the origin and another 
$w_B$ centred at $z=b\in \mathbb{C}$ 
\begin{equation}
w_{A}=\frac{i \kappa e^{-i\alpha}}{A \bar{z}}, 
\qquad w_{B}=\frac{ i \kappa e^{-i\alpha}}{A \left(\bar{z}-\bar{b}\right)} .
\label{eq:skyrmion_ansatz}
\end{equation}
The superposition is then given by 
\begin{equation}
w=\frac{i\kappa e^{-i\alpha}}{A}\frac{2\bar{z}
-\bar{b}}{\bar{z}\left(\bar{z}-\bar{b}\right)} .
\end{equation}
This has degree $Q=-2$ for all values of $b$, except $b=0$. 
At $b=0$ there is a cancellation between the numerator and 
the denominator and we end up with 
$w=\frac{2i\kappa e^{-i\alpha}}{A\bar{z}}$,
a $Q=-1$ configuration that is not a solution of the 
equations of motion. 

We define the interaction energy of the pair of skyrmions 
$w_{A},w_{B}$ to be
\begin{equation}
E_{\text{int}}[w]=E[w]-E[w_{A}]-E[w_{B}], \label{interaction energy}
\end{equation}
and regard it as the potential energy of the two-body force 
between the skyrmion pair. 
Our energy functional $E$ is dimensionless (a dimensionful 
constant is divided out from the physical energy). 
With this convention, the interaction energy $E_{\text{int}}[w]$ 
is also dimensionless. 
For dimensional reasons, $E_{\text{int}}[w]$ is a function of only two 
dimensionless ratios of the three dimensionful parameters, 
$\kappa, A, b$. 
We define dimensionless coordinates $\zeta, \beta$ as 
\begin{equation}
\zeta=\frac{2A}{\kappa}e^{-i\alpha}z, \quad 
\beta=\frac{2A}{\kappa}e^{-i\alpha}b. 
\label{eq:dimensionless-coordinate}
\end{equation}
The physical (dimensionful) separation $|b|$ between two skyrmions 
is given in terms of the dimensionless separation $|\beta|$ as 
$|b|=|\beta| r/2$. 
In Appendix \ref{interaction energy density computation}, we 
show that the interaction energy has the following 
simple structure as a function of two dimensionless parameters 
$\beta$ and $\kappa^2/(2A)$ 

\begin{equation}
E_{\text{int}}[w]
=\left[\left(\mathcal{I}_{2}^{\text{sup}}
\left(|\beta|\right) +16\pi\right) \right]\frac{\kappa^{2}}{2A} , 
\label{interaction-energy-dependence}
\end{equation}

\begin{equation}
\mathcal{I}_{2}^{\text{sup}}\left(|\beta|\right)
=\int 2\left(\frac{\vert w\vert^{4}
+4\text{Im}\left(w^{2}\partial_{\zeta}\bar{w}\right)}
{\left(1+|w|^{2}\right)^{2}}\right)d^{2}\zeta .
\label{I2-superposition}
\end{equation}
We observe that the interaction energy is independent of the angle 
parameter $\alpha$ of the DM interaction, and depends on 
$\kappa^2/(2A)$ only linearly. The coefficient 
$\mathcal{I}_{2}^{\text{sup}}$ depends 
only on the magnitude of the dimensionless separation, $|\beta|$.

\subsection{Qualitative analysis of interactions}

At the BPS point $B=2A=\kappa^2$, the interaction 
energy of the superposition $w$ is positive. This is demonstrated 
in the following way. 
The completing the square argument from Ref.~\cite{BRS} can be 
applied to Eq.~\eqref{energy functional-w} with $A=\kappa^2/2$. 
Using the superposition $w$ with the degree $Q=-2$, we obtain 
\begin{equation}
E[w]=-8\pi +8\int
 \frac{\left|\partial_{\bar z}w-\frac{i}{2}\kappa e^{i\alpha}w^2\right|^2}
{(1+|w|^2)^2} d^{2}x \geq -8\pi.
\end{equation} 
The second term is zero when the BPS 
equations are 
satisfied and is positive for every other configuration. 
This means that the energy in the degree $Q=-2$ sector is bounded 
below by $-8\pi$. Now from the definition of the interaction 
energy in \eqref{interaction energy} we have 
\begin{equation}
E_{\text{int}}[w]=E[w]-E[w_{A}]-E[w_{B}]
=E[w]+8\pi\geq 0,
\end{equation}
where we have used that both $w_{A}$ and $w_{B}$ 
are BPS skyrmions with energy $E=-4\pi$. 
This means that the interaction energy, $E_{\text{int}}$ must 
be positive to respect the completing the square argument. 
We also know that the interaction energy must vanish in the 
limit of infinite separation. 
Combining these facts, we find that the interaction is 
repulsive at least for asymptotically large separations. 

It was not known if the interaction energy is positive for other 
values of $\frac{A}{\kappa^2}$. 
For the range of $A$ which we consider, we find 
that the interaction energy between the exact skyrmion solutions 
is always positive. 

 For general external magnetic fields and anisotropies there 
are no known exact solutions so a different approach is needed 
to study the interaction energy. One way to approach this is 
to consider the asymptotic form of the equation for the hedgehog 
profile, $\Theta(r)$, as is done for $A=0$ in \cite{FKATDS,BLH}. 
Another approach used in \cite{CGC}, again for $A=0$, is to 
proceed numerically. Both of these approaches find that the 
interaction energy is repulsive and decreases exponentially 
for increasing separation.

%%%%
\subsection{Computation of the interaction energy}

We are now ready to compute the interaction energy 
between two exact single skyrmion solutions $w_{A}$ and $w_{B}$ 
in Eq.~\eqref{eq:skyrmion_ansatz}. 
Our approach proceeds by evaluating Eq.~\eqref{interaction-energy-dependence} 
for the superposition $w=w_A+w_B$.  
We can assume without loss of generality that $\beta$ is a 
real parameter. This is because the invariance of the energy under rotations, 
$w(\xi)\to e^{i\theta}w\left(e^{-i\theta}\xi\right)$, means 
that the interaction energy just depends on the modulus of the 
dimensionless separation, $\vert\beta\vert$. 
We compute the integrand of Eq.~\eqref{I2-superposition} and 
numerically integrate it using Mathematica, to find an expression 
for $E_{\text{int}}[w]$ using Eq.~\eqref{interaction-energy-dependence}. 
The integration is carried out in a $2\times10^{6}$ by $2\times 10^{6}$ 
region so that we can approximate the boundary as being at 
infinity compared to the skyrmion separation. After repeating the 
integration at many separations, between $|\beta|=30$ and $|\beta|=500$ 
we fit the interaction energy as a function of the dimensionless 
separation $|\beta|$ using a log-log plot
and find:
\begin{equation}
\mathcal{I}_{2}^{\text{sup}}\left(|\beta|\right) +16\pi
\simeq\frac{1592}{|\beta|^{1.735}}.
\end{equation}
The interaction energy is then
\begin{equation}
E_{\text{int}}[w]\simeq\frac{\kappa^2}{2A}\frac{1592}{|\beta|^{1.735}}.
\label{eq:interaction-energy}
\end{equation}

\begin{figure}
\begin{center}
\includegraphics[width=0.45\textwidth]{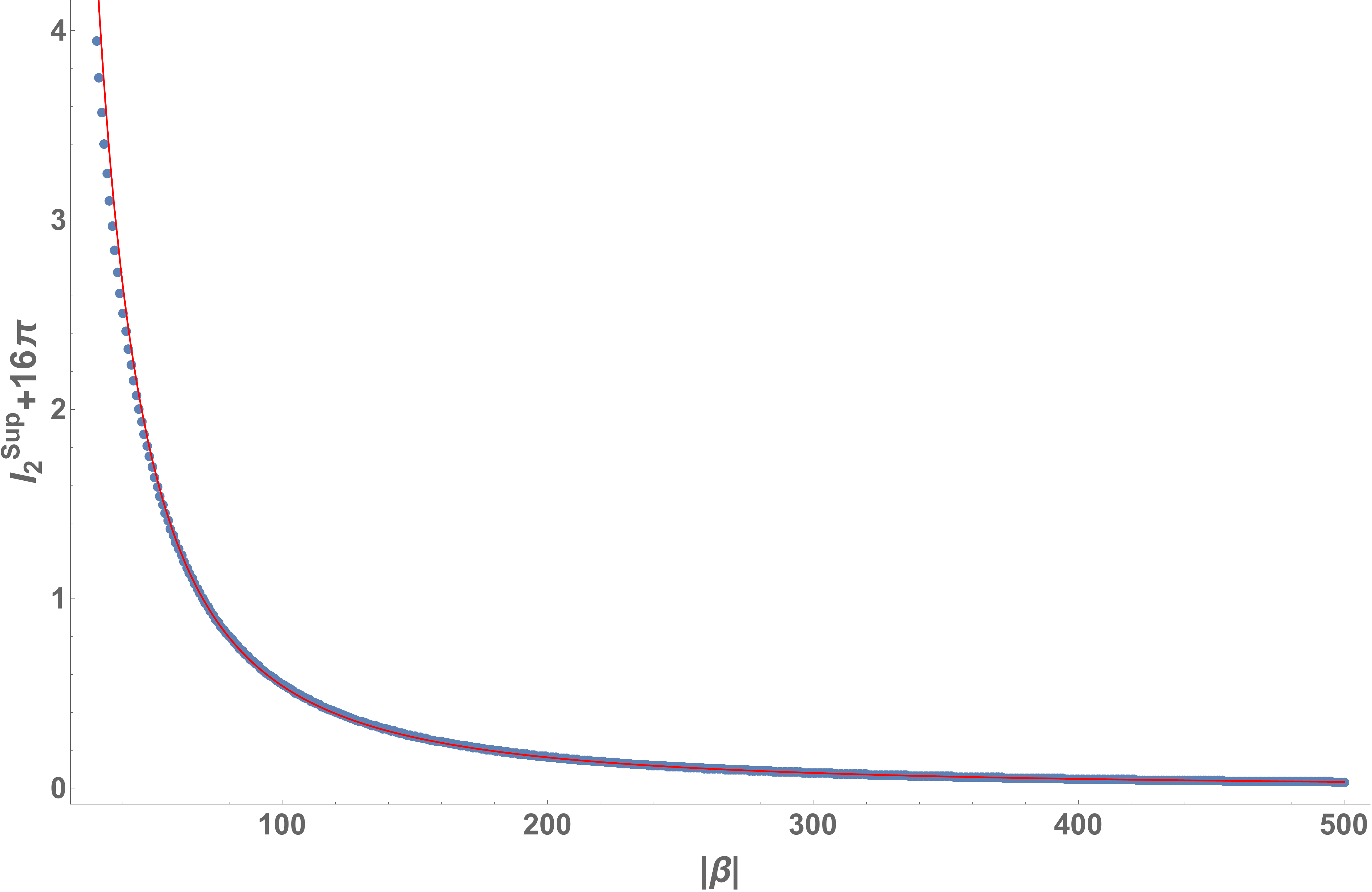}
\includegraphics[width=0.45\textwidth]{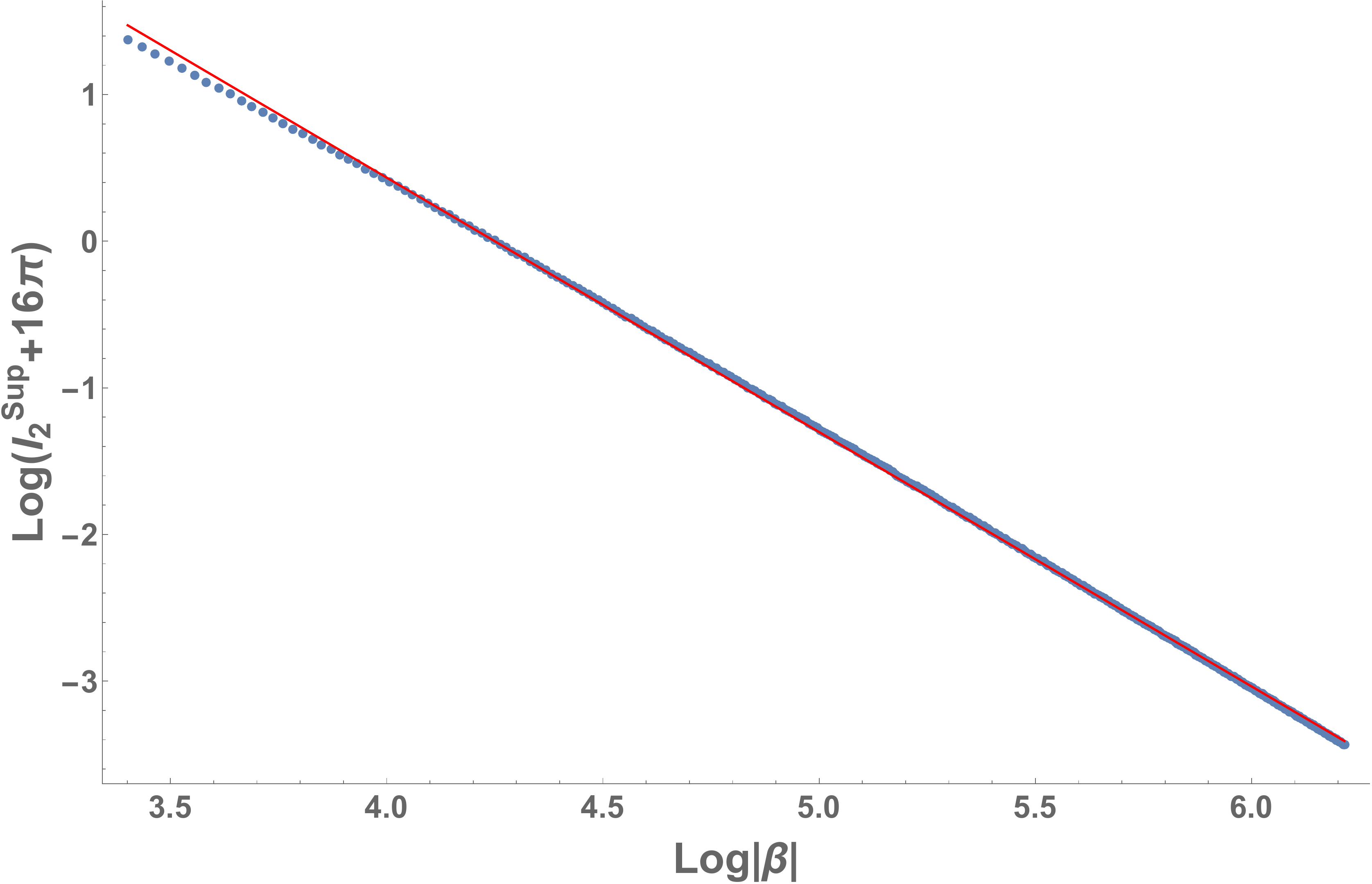}
\caption{These are fitted plots of $\mathcal{I}_{2}^{\text{sup}}
\left(|\beta|\right) +16\pi$ for $30<|\beta|<500$. On the left 
$\mathcal{I}_{2}^{\text{sup}}\left(|\beta|\right) +16\pi$ is 
plotted against the dimensionless separation $|\beta|$, while on 
the right $\log\left(\mathcal{I}_{2}^{\text{sup}}\left(|\beta|\right) 
+16\pi\right)$ is plotted against $\log\vert\beta\vert$. The 
data points are the numerically computed values of 
$\mathcal{I}_{2}^{\text{sup}}+16\pi$ and the red fit line is 
$\frac{1592}{|\beta|^{1.735}}$.} 
\label{critical coupling interaction fig}
\end{center}
\end{figure}

The power law fit for $\mathcal{I}_{2}^{\text{sup}}\left(|\beta|\right) 
+16\pi$ is plotted in Fig.~\ref{critical coupling interaction fig}. 
There is good agreement between the fit and the numerical computation 
for larger separations. At small separations there is a slight 
deviation with Eq.~\eqref{eq:interaction-energy} marginally over 
estimating the strength of the interaction.

The general trends that we observe are that skyrmions are 
small, weakly interacting, and have positive energy for large values 
of $\frac{A}{\kappa^{2}}$. As $A$ decreases towards $\kappa^{2}$, 
the inter-skyrmion interaction 
gets stronger and the skyrmions increase in size and decrease in 
energy, with the energy of an individual skyrmion vanishing at 
$A=\kappa^{2}$, the phase transition point between the ferromagnetic 
and skyrmion lattice phases. 

The interaction gets stronger and the skyrmion energy 
becomes more negative as $A$ decreases. 
On the other hand, the relative separation $|\beta|$ becomes 
smaller as $\frac{A}{\kappa^{2}}$ decreases. 
Therefore our results get less reliable as $A$ decreases, 
and it is likely that we are overestimating the repulsive 
interaction energy. This can be observed directly from 
Fig.~\ref{superposition magnetisation vector fig} where for 
smaller values of $A$ the distinction between the two single 
skyrmions becomes less pronounced. At short distances, the 
skyrmions will have a significant effect on each others shape 
and a more sophisticated method is needed to account for this.

%%%%%%%%%%%%%%%%%%%%%
\section{Skyrmion lattice states}\label{sec:lattice}

\subsection{Average energy density of skyrmion lattice state}
For $A<\kappa^2$, the homogeneous ferromagnetic state 
becomes unstable due to the emergence of negative energy single 
skyrmion solutions. 
The negative energy of individual skyrmions results in a preference 
to produce as many skyrmions as possible. 
However, the repulsive interaction prevents the skyrmions 
from becoming too densely packed. 
This competition is expected to lead to a lattice of skyrmions.
Depending on symmetries, either a triangular or square lattice 
configuration will have the lowest energy. An extensive numerical 
study in Ref.~\cite{LSB} indicated that a triangular 
lattice configuration is the ground state for the parameter range 
where the exact skyrmion solutions occur, 
as in the case of Abrikosov lattices of vortices in superconductors and superfluids. 
A cubic lattice of merons, half skyrmions, was also found numerically 
in another parameter region with larger anisotropy, $B<2A$. 
In light of this numerical work we construct a simple model 
for a triangular lattice of skyrmions. Using this model, we 
compute the average energy per unit cell and find the lattice 
spacing which minimises this.

Assuming that the two-body force between skyrmions is 
the dominant interaction, we estimate the average energy density 
of the triangular lattice state. 
Let us identify the lattice spacing to be the separation $|b|$ 
in the calculation of the two-body force between skyrmions in 
Eq.\eqref{eq:skyrmion_ansatz}. 
We find it convenient to use a dimensionless separation $\ell$ 
defined as the ratio of $|b|$ to the diameter of a single skyrmion 
$2r=2\kappa/A$ 
\begin{equation}
\ell=\frac{|b|}{2r}=\frac{|b|A}{2\kappa}=\frac{|\beta|}{4}. 
\label{eq:dimless-separation}
\end{equation}
Using the interaction energies computed in the previous 
section, we now compute the approximate energy density 
for a triangular skyrmion lattice.  
\begin{figure}[htbp]
\begin{center}
\includegraphics[width=0.3\textwidth]{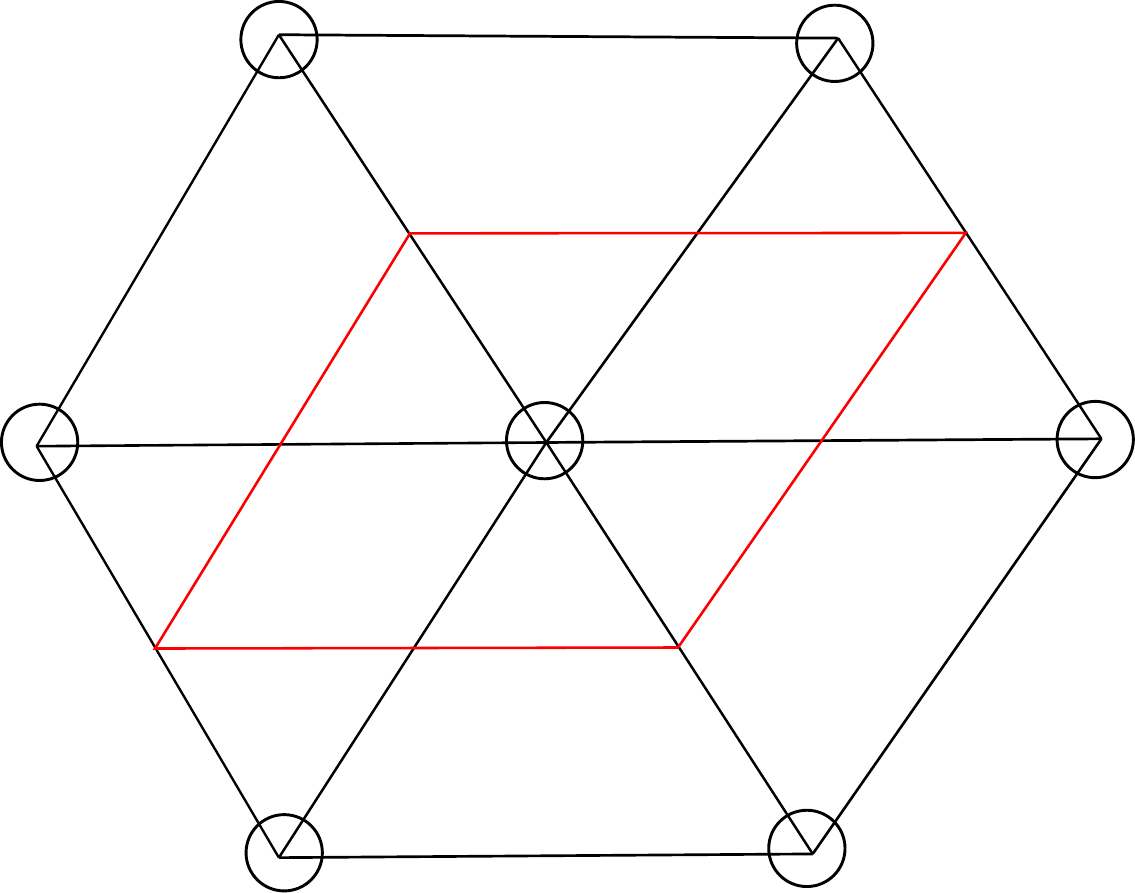}
\caption{The unit cell of the triangular lattice is the red parallelogram.}
\label{unit cell fig}
\end{center}
\end{figure}
Our approach is to compute the energy per unit cell. 
The unit cell contains one skyrmion, with 6 
nearest neighbours, and has area
$\frac{\sqrt{3}}{2}|b|^{2}=\frac{\sqrt{3}}{2}(\frac{2\kappa\ell}{A})^{2}$, 
as shown in Fig.~\ref{unit cell fig}. 
The average energy density 
is then given by 
\begin{equation}
\langle e_{\text{Lattice}} \rangle\left(\ell\right) =\left(\frac{A}{2\kappa}\right)^{2}\frac{2}{\sqrt{3}\ell^{2}}\left(E_{\text{Sky}}+3E_{\text{int}}\right)
=\left(\frac{A}{2\kappa}\right)^{2}\frac{2}{\sqrt{3}\ell^{2}}\left(E_{\text{Sky}}+3\frac{\kappa^{2}C_{1}}{2A \ell^{1.735}}\right),
\label{energy per triangular unit cell}
\end{equation}
where $E_{\text{Sky}}$ is the skyrmion energy given in 
Eq.~\eqref{hedgehog energy expression}, and 
$C_{1}\simeq\frac{1592}{4^{1.735}}$ 
is the constant given in Eq.~\eqref{eq:interaction-energy} 
when substituting Eq.~\eqref{eq:dimless-separation} for $|\beta|$.

\begin{figure}[htbp]
\begin{center}
\includegraphics[width=0.45\textwidth]{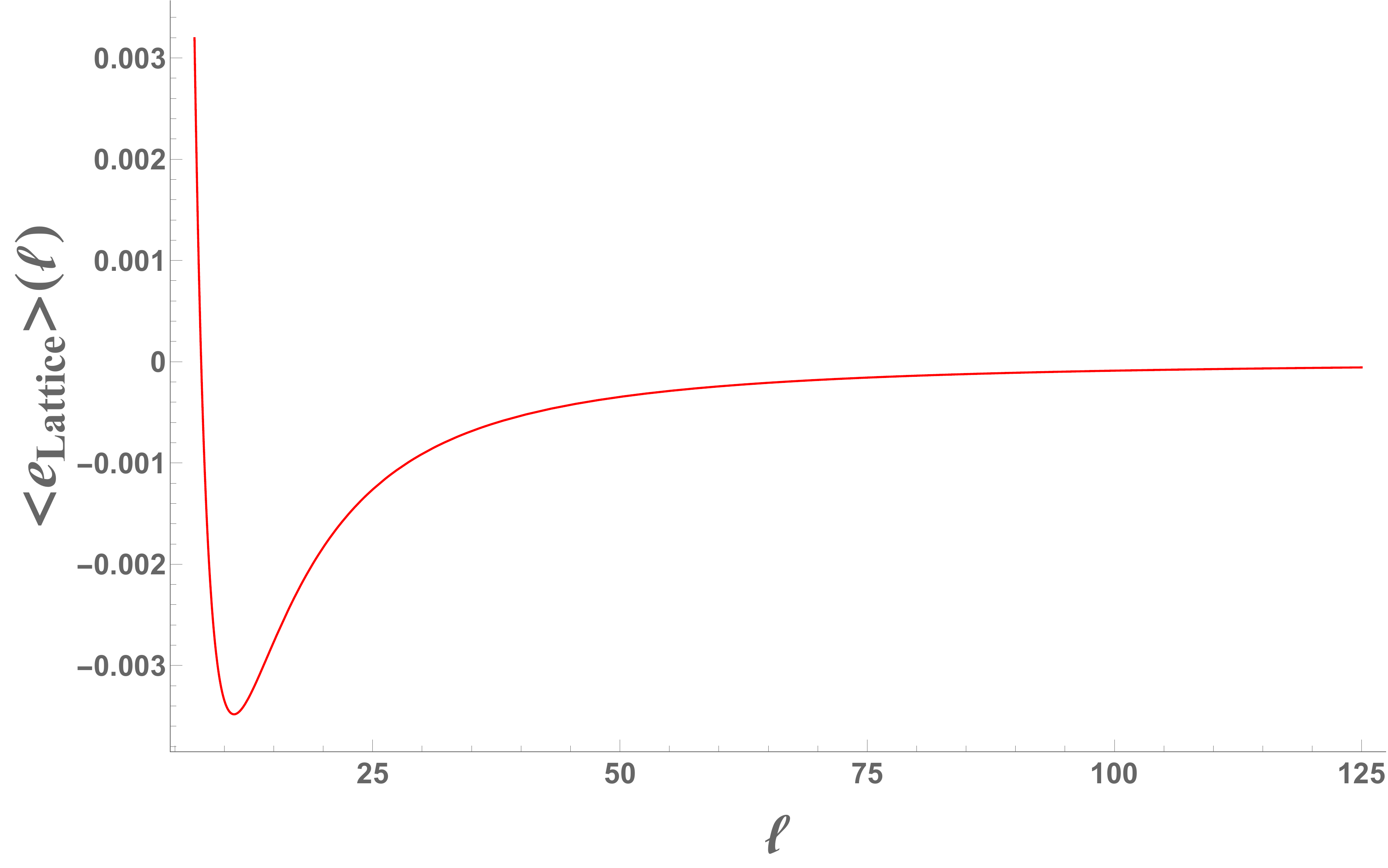}
\includegraphics[width=0.45\textwidth]{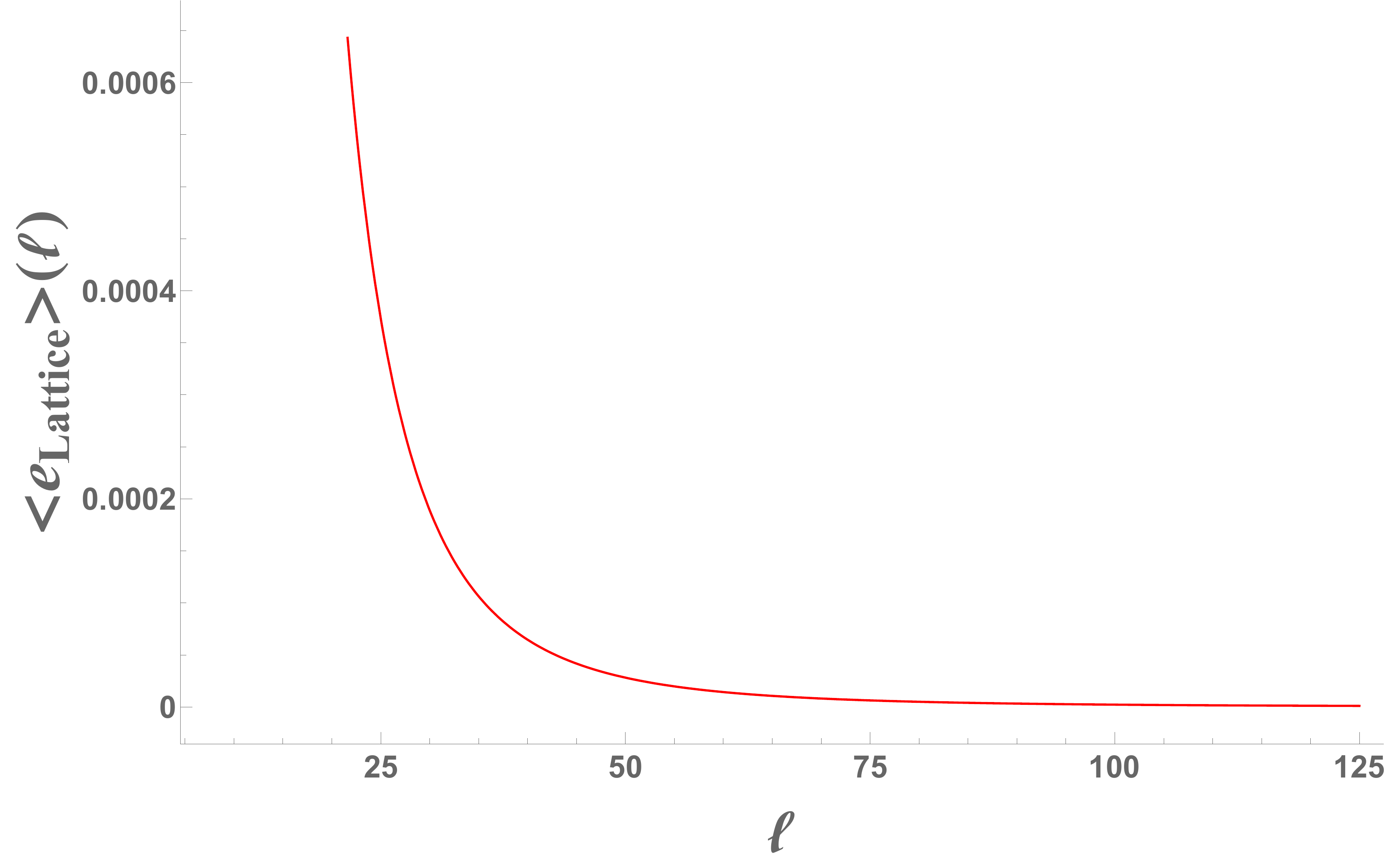}
\caption{These are plots of the energy per unit cell from Eq.~\eqref{energy per triangular unit cell}. On the left is the critically coupled case $A=\frac{\kappa^{2}}{2}$ with its minimum at $\ell\simeq 11$ and on the right is $A=\kappa^{2}$ whose minimum is for infinitely separated skyrmions.}
\label{lattice energy fig}
\end{center}
\end{figure}

Depending on the value of $\frac{A}{\kappa^{2}}$, 
Eq.~\eqref{energy per triangular unit cell} will have different 
behaviour as a function of $\ell$. In Fig.~\ref{lattice energy fig} 
we plot $\langle e_{\text{Lattice}} \rangle$ as a function of 
$\ell$ for the BPS case, $\frac{A}{\kappa^{2}}=\frac{1}{2}$, and 
at the phase transition to the ferromagnetic phase, 
$\frac{A}{\kappa^{2}}=1$. In the BPS case $\langle e_{\text{Lattice}} 
\rangle$ is negative at its minimum and there the triangular 
skyrmion lattice is the ground state. While for 
$\frac{A}{\kappa^{2}}\geq1$ the minimum of 
$\langle e_{\text{Lattice}} \rangle$ is positive and reached for 
asymptotically separated skyrmions. This is because the ferromagnetic 
state is the ground state and skyrmions arise as positive energy 
excitations above it.

When $\frac{A}{\kappa^{2}}<1$  minimising Eq.~\eqref{energy per triangular unit cell} with respect to $\ell$ leads to $\ell=\ell_{\text{min}}$ with
\begin{equation}
\ell_{\text{min}}\simeq\left(-\frac{\left(2+\delta\right)3C_{1}}{16\pi\left(\frac{A}{\kappa^{2}}-1\right)}\right)^{\frac{1}{\delta}}, \label{lattice spacing}
\end{equation}
where $\delta=1.735$.

\begin{figure}[htbp]
\begin{center}
\includegraphics[width=0.45\textwidth]{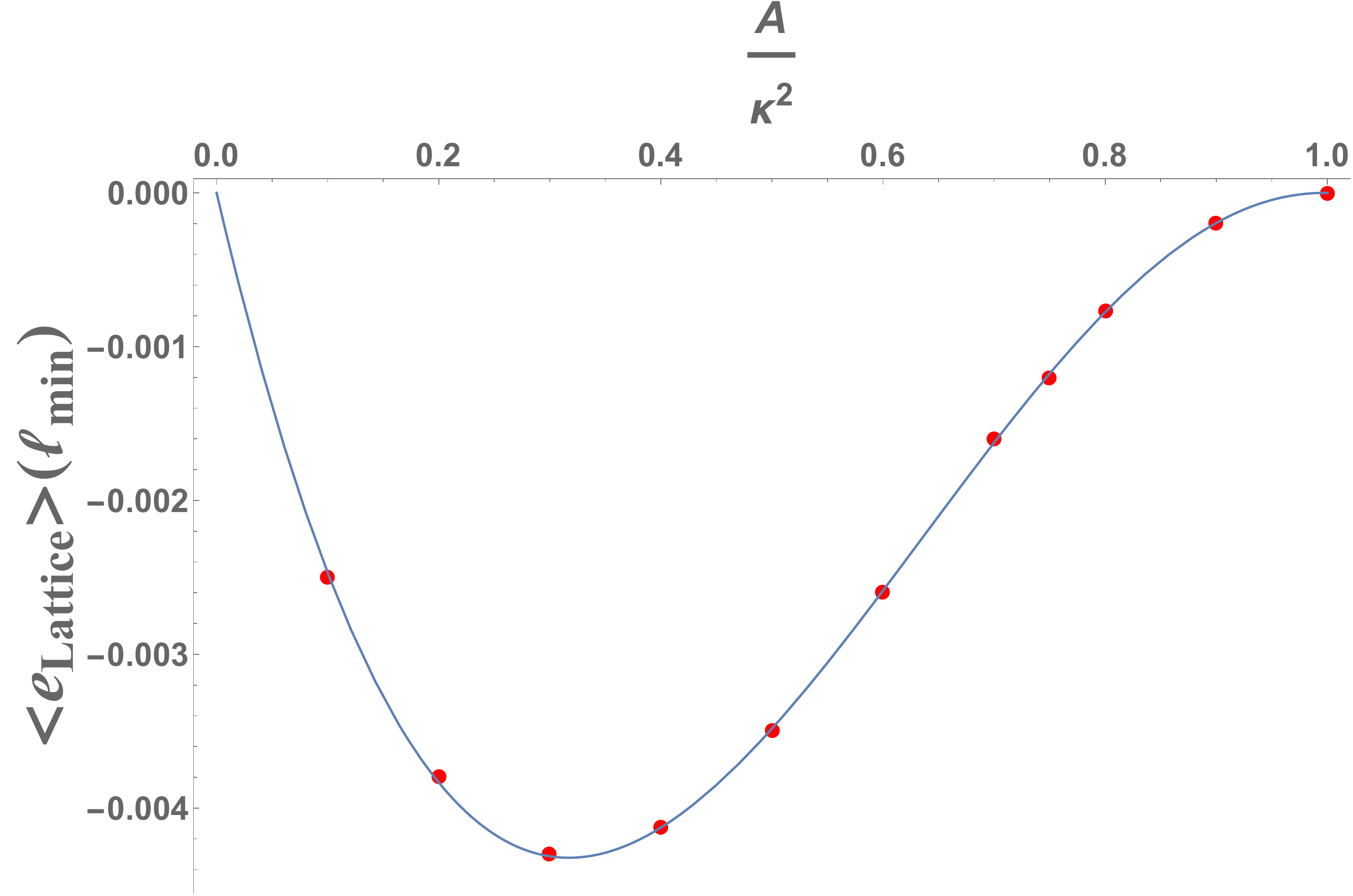}
\caption{This figure shows how the average energy density of the lattice, \eqref{energy per triangular unit cell}, at the minimum changes with $A$. The blue line is found by substituting the approximate value of $\ell_{\text{min}}$ from Eq.~\eqref{lattice spacing} into Eq.~\eqref{energy per triangular unit cell} and the red points are found numerically. }
\label{lattice energy density fig}
\end{center}
\end{figure}

\begin{figure}[htbp]
\begin{center}
\includegraphics[width=0.45\textwidth]{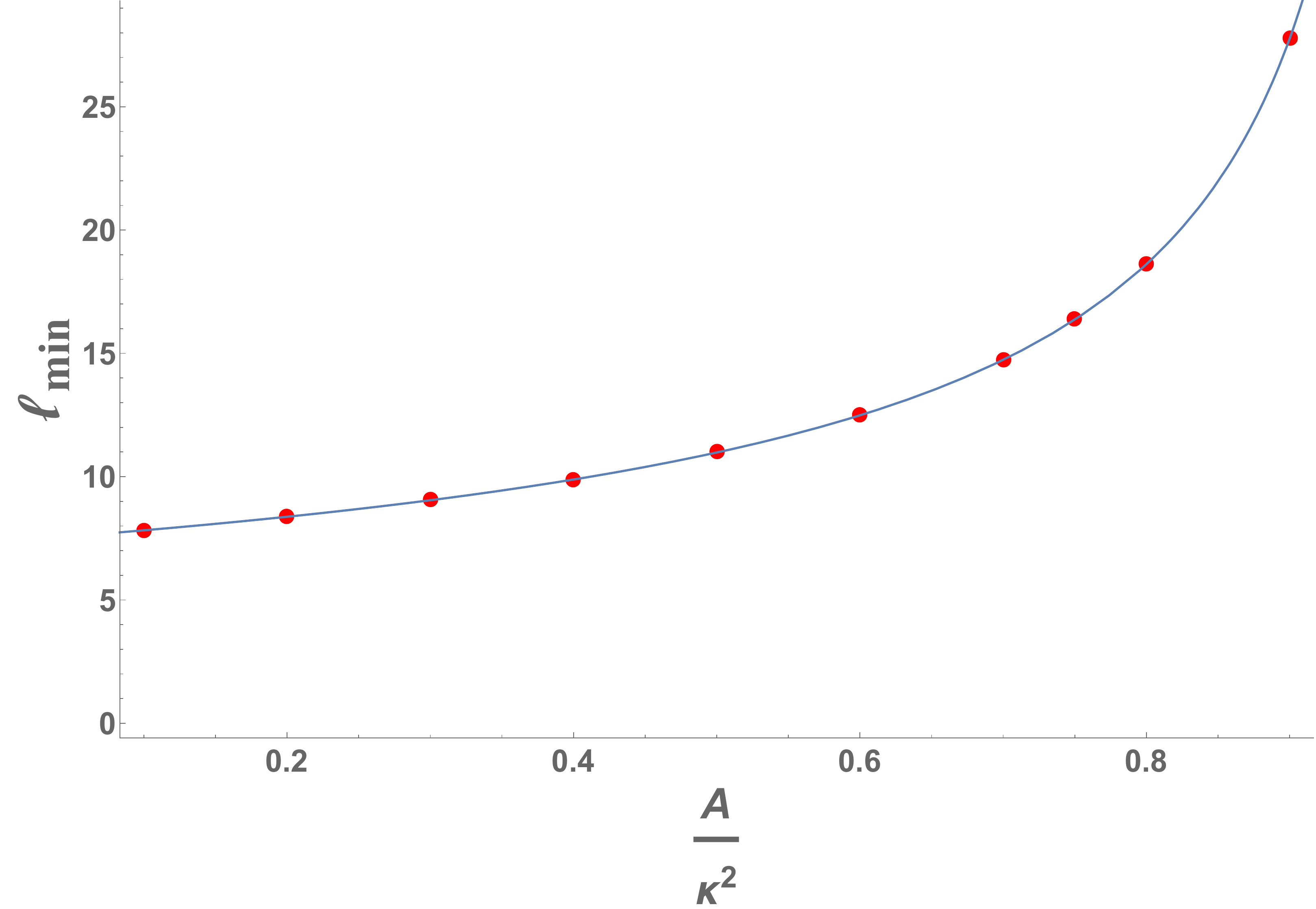}
\caption{Here we plot the dimensionless lattice separation $\ell_{\text{min}}$ from Eq.~\eqref{lattice spacing} against $\frac{A}{\kappa^{2}}$. As $A$ decreases $\ell_{\text{min}}$ also decreases. The blue curve is a plot of the approximate value of $\ell_{\text{min}}$ from Eq.~\eqref{lattice spacing} and the red points are the numerically computed values of $\ell_{\text{min}}$. }
\label{lattice spacing fig}
\end{center}
\end{figure}

The average energy density $\langle e_{\text{Lattice}} \rangle\left(\ell_{\text{min}}\right)$ 
and the dimensionless lattice spacing at the minimum, $\ell_{\text{min}}$, are plotted against 
$\frac{A}{\kappa^{2}}$ in Fig.~\ref{lattice energy density fig} and 
Fig.~\ref{lattice spacing fig}, respectively. 
These figures show that as $\frac{A}{\kappa^{2}}$ decreases the dimensionless lattice separation at the minimum
decreases. While at first the lattice energy density decreases with decreasing $\frac{A}{\kappa^{2}}$ it reaches a minimum near $\frac{A}{\kappa^{2}}=0.3$ before increasing towards zero.

Let us make a few observations. 
Firstly we observe that a finite lattice spacing is achieved 
by the balance between the negative energy of constituent skyrmions 
and the positive energy due to the repulsive interaction between 
skyrmions. Secondly, this semi-analytic understanding is most reliable 
quantitatively 
for regions near the phase transition point $\frac{A}{\kappa^{2}}=1$. 
Thirdly for small $\frac{A}{\kappa^{2}}$, $|\beta|$ becomes small.
Hence our calculation of the interaction energy becomes less reliable. 
Since we are using a superposition of single skyrmion solutions 
as a variational Ansatz for a $Q=-2$ configuration, we naturally 
expect that the true total energy at a fixed separation, for a 
$Q=-2$ configuration, may be lower. 
Namely it is quite likely that our approximate result for the interaction 
energy is overestimating the repulsive interaction.
Therefore we expect that the skyrmion separation in the lattice will be smaller than our estimate for small values of $\frac{A}{\kappa^{2}}$.

\begin{figure}[htbp]
\begin{center}
\includegraphics[width=0.45\textwidth]{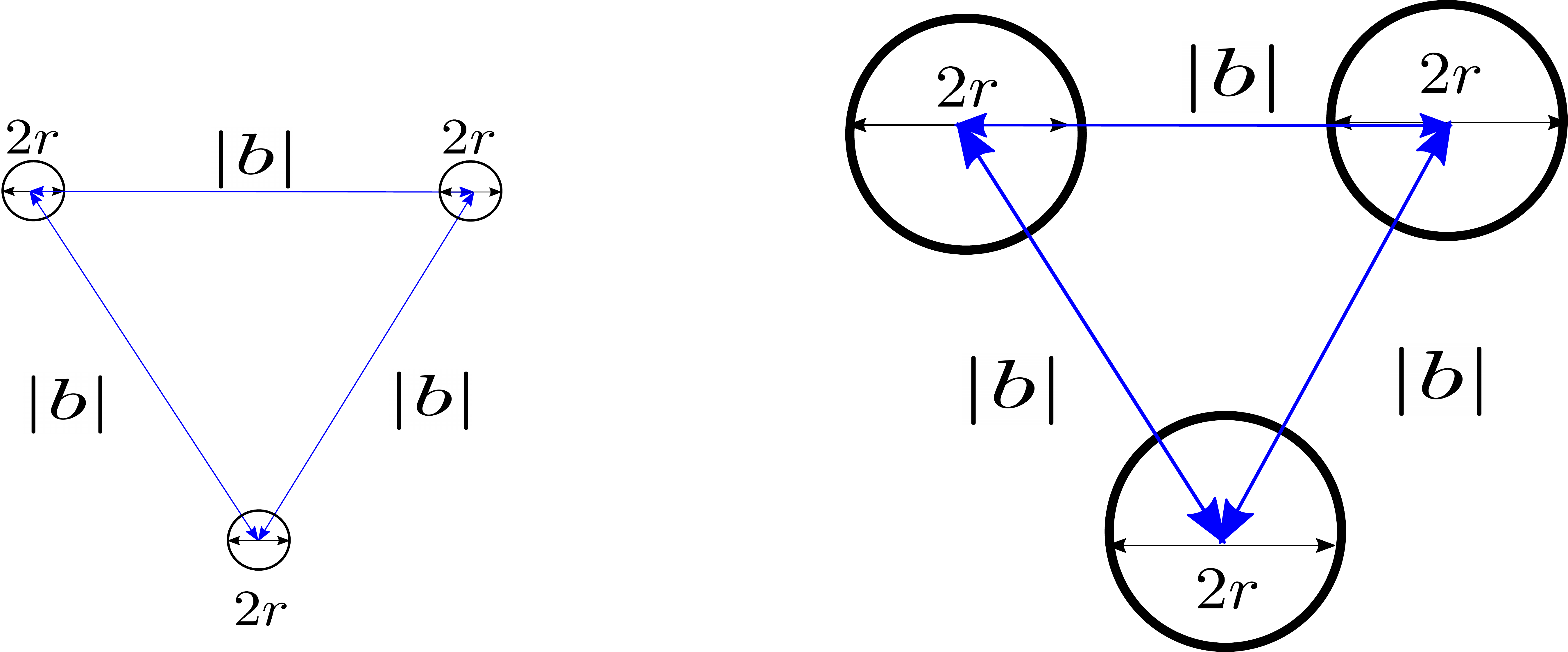}
\caption{A schematic of showing how the skyrmion diameter $2r$ 
and the lattice separation $|b|$ change with decreasing $\frac{A}{\kappa^{2}}$. 
On the left for larger $A$ the lattice spacing, $\vert b\vert$ 
is much larger than the diameter of the skyrmion, $2r$. 
On the right for lower $\frac{A}{\kappa^{2}}$ both the skyrmion diameter 
and lattice separation increase but their ratio, 
$\ell=\frac{\vert b\vert}{2r}$, decreases. The scale in the figure on the right has been exaggerated for effect.}
\label{sls}
\end{center}
\end{figure}

\subsection{Transition to the spiral state}

For smaller values of the parameter $A$, compared to the 
DM interaction parameter $\kappa^2$, we expect from \cite{LSB} that spiral 
states have lower energy than the lattice. Thus we expect that a phase transition between the skyrmion 
lattice and spiral states occurs. 
As we have seen in the preceding section, our estimate of 
interaction energy is likely to be an overestimate for smaller 
values of $\frac{A}{\kappa^{2}}$.
On the other hand, the exact single skyrmion solution has energy 
$E_{\rm Sky}=4\pi(1-\frac{\kappa^2}{A})\to-\infty$ and radius 
$r=\frac{\kappa}{A}\to\infty$ as $A\to 0$, with the average energy density 
$\langle e_{\text{Sk}}\rangle$ inside the radius tending 
to a finite value. 

From Fig.~\ref{lattice spacing fig} we see that for small $\frac{A}{\kappa^{2}}$ the lattice spacing and the skyrmion diameter become the same order of magnitude. This suggests that near the transition between the lattice and spiral phases we need a better model of the lattice. 
As a candidate model of this kind, we consider the case of skyrmions separated by a multiple of the skyrmion diameter,
\begin{equation}
|b|=2c\frac{\kappa}{A} \label{lattice spacing small A}
\end{equation}
with $c$ a positive constant. The case of $c=1$ would have skyrmions separated by $2\frac{\kappa}{A}$, the skyrmion diameter, which means that the magnetisation vector $\vec{n}$ will not attain the value $e_{3}$ between the skyrmions. A situation like this is called a hexagonal meron lattice in \cite{SKC}. As such we are interested in $c>1$ and will in fact focus on $c=2$. We are also assuming that the repulsive skyrmion interaction can be neglected, except to keep the skyrmions from merging, such that they form a lattice. 
This leads to the situation depicted on the right in Fig.~\ref{sls}. This type of triangular lattice made from nearly touching skyrmions was considered in \cite{HZZJN}. This scenario seems to be more realistic for small values of $\frac{A}{\kappa^{2}}$, since we are overestimating the interaction energy in that region.

Using this alternative model of a skyrmion lattice we can 
estimate the value of $A$ that the transition between the lattice 
phase to the spiral phase occurs at.
To do this we note that for $V(n_{3})=0$ (at $A=B=0$)
there is an exact spiral solution 
\begin{equation}
\vec{n}=\begin{pmatrix}
0\\
-\sin\left(\kappa x\right)\\
\phantom{-}\cos\left(\kappa x\right)
\end{pmatrix} \label{average spiral energy}
\end{equation}
with average energy
\begin{equation}
\langle e_{\text{spiral}}\rangle =-\frac{\kappa^{2}}{2}.
\end{equation}
The most general spiral solution at $A=B=0$ is obtained 
by rotating and/or translating this solution in the $x$-$y$ plane with 
the simultaneous rotation of the magnetization vector $\vec{n}$.
This has an identical energy density so we choose to work with the simplest case, 
Eq.~\eqref{average spiral energy}. 
When the potential is non-zero the spiral configuration given 
by Eq.~\eqref{average spiral energy} has average energy 
\begin{equation}
\langle e_{\text{spiral}}\rangle =-\frac{\kappa^{2}}{2}+\frac{3A}{2},
\end{equation}
which is shown in Appendix \ref{spiral state appendix}. 
While it is not a solution of the equations of motion it provides 
an approximation for a spiral state when the potential is small. In \cite{Melcher} for $A=0, B\neq 0$ a spiral state like Eq.~\eqref{average spiral energy} was motivated by comparing the energy functional Eq.~\eqref{energy-functional-general} to an energy functional whose ground states are Beltrami fields, solutions to $\nabla \times \vec{n}+\kappa\vec{n}=0$. 
A rough estimate of where the transition between the skyrmion 
lattice and the spiral takes place can be found by comparing 
this to the average energy density of a single skyrmion. 
The skyrmion average energy is given by  Eq.~\eqref{average skyrmion energy density general radius}.
The transition will occur when these two energy densities are 
equal. When the separation is $|b|=4\frac{\kappa}{A}$ (taking $c=2$ in Eq.~\eqref{lattice spacing small A}) this gives the phase transition at 
\begin{equation}
A=\frac{\kappa^{2}}{16}\left(20-4\sqrt{17}\right)\simeq 0.22\kappa^{2},
\end{equation}
this value of $A$ matches the location of the phase transition found numerically in \cite{LSB}.
We refer the reader to  Appendix \ref{spiral state appendix} for the explicit details of this computation. 
At this value of the coupling the spiral state has energy density 
\begin{equation}
\langle e_{\text{spiral}}\rangle\simeq -0.17\kappa^{2}.
\end{equation}
This energy density is much lower than the energy density 
that we found for the skyrmion lattice in Fig.~\ref{lattice energy density fig} 
using the dipole approximation.
This is further evidence that our dipole approximation underestimates 
the energy density because it is over estimating the interaction energy.

\section{Conclusion}\label{sec:summary}

In this paper, we have studied the interaction between two 
well-separated skyrmions and the phase structure of a chiral magnet. 
We have focused on the solvable line $B=2A$, where $A$ is the anisotorpy 
parameter and $B$ is the magnetic field parameter in the potential,  so that
we can make use of the exact skyrmion configurations from Ref.~\cite{BRS}. 
For $\frac{A}{\kappa^2}\gg 1$, where the DM interaction can be neglected 
compared to the potential, we have found the homogeneous ferromagnetic 
state $n_3=1$ as the ground state. 
This ferromagnetic phase continues for smaller values of 
$\frac{A}{\kappa^2}$ until we reach the critical point at 
$\frac{A}{\kappa^2}=1$, where the exact skyrmion solution has zero energy. 
Below this point $\frac{A}{\kappa^2}<1$, the exact skyrmion 
solution has negative energy. The presence of a negative energy solution to the equations of motion means that the homogeneous ferromagnetic state is no longer the lowest energy state.  This suggests that the true ground state becomes an inhomogeneous skyrmion lattice state as is found numerically in Ref.~\cite{LSB,APL_thinfilms_2016,MRG_2016}. As the skyrmion energy is known analytically the transition point, $A=\kappa^{2}$, is found exactly without the ambiguity of numerical simulations. This is only possible on the solvable line where exact skyrmion solutions are known.

We have studied the properties of this skyrmion lattice state 
by working out the interaction between exact skyrmion 
configurations along the solvable line $B=2A$.
To achieve this we used the dipole approximation for the superposition of 
two, exact, single skyrmion solutions centred with a finite separation. 
and have found that the interaction energy is independent of the angle 
parameter of the DM interaction which distinguishes Bloch, N\'{e}el 
or other types of skyrmion.
Exact expressions for the interaction energy density were obtained 
and after numerically integrating these a power law fit is found 
for the interaction energy of well separated skyrmions. 
Using the computed interaction energies, we have been able to predict 
that the phase transition from a skyrmion lattice state to a 
ferromagnetic state will happen for $\frac{A}{\kappa^2}=1$. 
Our results have also demonstrated that in the skyrmion lattice 
phase as $\frac{A}{\kappa^2}$ decreases the ratio of lattice 
separation to the skyrmion diameter decreases. This suggests 
that near the boundary between the lattice and spiral phases 
the lattice can be approximately computed by taking the lattice 
density to be the energy density of a single skyrmion. 
This allows us to gain a semi-analytic understanding of the 
phase diagram of a chiral magnet along the solvable line. 
In particular we determine the location of the phase 
transition between the homogeneous ferromagnetic phase and 
the skyrmion lattice phase precisely, and estimated the transition 
between the skyrmion lattice phase and the spiral phase.

We have emphasized the need for the boundary term in the energy 
functional to make the DM interaction term well defined, and have 
clarified its close relation to the asymptotic behavior of the
magnetisation vector. 

We need to emphasise that our work deals only with static 
configurations, since the exact skyrmion configurations 
from Ref.~\cite{BRS} that we have used are minimisers of the 
static energy functional. 
Understanding the effect of time evolution of excited states above 
the ground state is an open problem, which requires an 
understanding of the dynamics of skyrmion configurations 
and their superposition. Another interesting open problem is to find the values of $A$ and $B$ at which the single skyrmion energy becomes negative away from the solvable line.
 
We have studied skyrmions in two dimensions, applicable to a thin film.
In three spatial dimensions, skyrmions are lines for which 
one can study collective modes propagating along a skyrmion line \cite{Kobayashi:2014eqa}.
Extension of the present work to three dimensions remains an open problem.

%%%%%%%%%%%%%%%%%%%%%%%%%%%%%%%%%%%%%%%%%%%%%%%%%%%%%
%%%%%%%%%%%%%%%%%%%%%%%%%%%%%%%%%%%%%%%%%%%%%%%%%%%%%
\begin{acknowledgments}
C.~R thanks Bernd Schroers and Bruno Barton-Singer for useful discussions about magnetic skyrmions and the importance of the boundary term, and Sven Bjarke Gudnason for general discussions about interacting solitons.
This work is supported by the Ministry of Education, Culture, 
Sports, Science, and Technology(MEXT)-Supported Program for the 
Strategic Research Foundation at Private Universities ``Topological 
Science" (Grant No. S1511006) 
and by the Japan Society for the Promotion of Science (JSPS) 
Grant-in-Aid for Scientific Research (KAKENHI) Grant Number
(16H03984 (M.~N.), 18H01217 (M.~N. and N.~S.)). 
The work of M.~N.~is also supported 
in part by a Grant-in-Aid for Scientific Research on Innovative Areas 
``Topological Materials Science'' 
(KAKENHI Grant No.~15H05855) from MEXT of Japan.

\end{acknowledgments}

%%%%%%%%%%%%%%%%%%%%%%%%%%%%%%%%%%%%%%%%%%%%%%%%%%%%%
%%%%%%%%%%%%%%%%%%%%%%%%%%%%%%%%%%%%%%%%%%%%%%%%%%%%%
\appendix
\section{DM interaction and boundary terms}
\label{boundary term discussion}
In this appendix, we explain why the inclusion of the boundary 
term in the energy functional is mandatory. 
Let us consider the energy functional $E[\vec{n}]_{\rm bulk}$ 
in the bulk with the DM interaction in 
Eq.~\eqref{energy-functional-general}. 
To obtain the equation of motion by the variational principle, 
we consider a variation of the bulk energy functional 
$\delta E[\vec{n}]_{\rm bulk}$ through the variation of the 
magnetization vector $\delta n_{\alpha}^a$ 
\begin{eqnarray}
\delta E[\vec{n}]_{\rm bulk}&=&\int%_{\mathbb{R}^{2}}
d^2 x \left[\delta n_a^{\alpha}\left(-\nabla^2 n_a^\alpha
+2\kappa\epsilon^{aib}\nabla_i n_b^\alpha
+\frac{\partial U}{\partial n_a^\alpha}\right)\right.
\nonumber \\
&+&\left.\nabla_i\left(\delta n_a^\alpha\left(\nabla_i n_a^\alpha
-\kappa \epsilon^{aib}n_b^\alpha\right)\right)\right] , 
\label{eq:variation_bulk_energy_density}
\end{eqnarray}
where we have performed a partial integration resulting in the 
total derivative term (the last term). 
We can obtain the equation of motion if and only if the total 
derivative term (surface term in the variation of the total energy) 
is absent. 
We can achieve this goal by adding a boundary term to the 
energy functional whose variation should satisfy 
\begin{eqnarray}
\delta E_{\rm boundary}=
-\int d^2 x \nabla_i\left(\delta n_a^\alpha
\left(\nabla_i n_a^{\alpha, {\rm bound}}
-\kappa \epsilon^{aib}n_b^{\alpha, {\rm bound}}\right)\right) , 
\label{eq:variation_boundary_energy_density}
\end{eqnarray}
where the booundry value of the magnetization vector is denoted 
as $n_a^{\alpha, {\rm bound}}$. 
We can take the following boundary energy functional as a 
solution of the above requirement 
\begin{eqnarray}
E_{\rm boundary}&=&
-\int d^2 x \nabla_i\left( n_a^\alpha\left(\nabla_i n_a^{\alpha, {\rm bound}}
-\kappa \epsilon^{aib}n_b^{\alpha, {\rm bound}}\right)\right) , 
\label{eq:boundary_energy_density}
\end{eqnarray}
where we denote the boundary value of the variable $n^\alpha$ 
as $n^{\alpha, {\rm bound}}$. 
When we evaluate the contribution from the first term, we note 
that $%\begin{equation}
n_a^\alpha%\left
(\nabla_i n_a^{\alpha, {\rm bound}}%\right
)\to n_a^{\alpha, {\rm bound}}%\left
(\nabla_i n_a^{\alpha, {\rm bound}}%\right
)=0$ %, \label{eq:kin_boundary_contr}\end{equation}
because of $(n_a^{\alpha, {\rm bound}})^2=1$. 
Therefore we need to consider only the second term. 
Thus the boundary energy functional is finally given by 
\begin{equation}
E_{\rm boundary}=\int d^2 x 
\nabla_i\left(
\kappa \epsilon^{aib} n_a^\alpha n_b^{\alpha, {\rm bound}}\right) . 
\label{eq:boundary_contr}
\end{equation}

For definiteness, let us consider the case of single skyrmion 
($Q=-1$) solution in the parameter region $2A \le B$. 
Since the minimum of the potential is at 
$\vec{n}^\alpha=(0,0,1)\equiv\vec{n}_\infty$, we should impose a 
boundary condition fixing the boundary value as 
$n_a^{\alpha, {\rm bound}}=n_{a, \infty}$. 
It is convenient to rewrite the boundary energy functional in 
Eq.~\eqref{eq:boundary_contr} using an identity in 
Ref.~\cite{BRS} 
\begin{eqnarray}
{E}_{\rm boundary}
%&
=%&
-\int dx_1\wedge dx_2 \kappa(\partial_1n_2^\alpha
-\partial_2n_1^\alpha)
%\nonumber \\&
=%&
-\int dx_1\wedge dx_2 \; \omega
 , 
\label{eq:DM_boundary_contr}
\end{eqnarray}
where $\omega$ stands for the vorticity density 
in the $n_1, n_2$ components. 
The vorticity $\Omega$ is defined as 
\begin{equation}
\Omega =\frac{1}{4\pi }\int_{\mathbb{R}^{2}} d^2 x \; \omega \; 
=\frac{1}{4\pi } \int_{\mathbb{R}^{2}}d^{2}x \;  \kappa \; 
\vec{n}_{\infty}\cdot \left(\nabla_{-\alpha}\times \vec{n}\right)
, 
\label{vorticity}
\end{equation}
in order to make variational principle well-defined. 
Thus we find that the correct energy of the skyrmion solution 
for $2A\le B$ is given by 
\begin{eqnarray}
\label{energy-functional}
E[\vec{n}]&=&E_{\rm bulk}-4\pi\Omega 
\\
&=&\int_{\mathbb{R}^{2}} \left[\frac{1}{2}|\nabla \vec{n}|^{2}
+\kappa\;\left(
\vec{n}-e_{3}\right)
\cdot \left(\nabla_{-\alpha}\times 
\vec{n}\right)+U(n_3)\right]d^{2}x . 
\end{eqnarray}
By choosing the potential $V(n_3)$ for the solvable line 
($B=2A$) instead of the generic $U(n_3)$, we obtain our energy 
functional in Eq.~\eqref{energy functional static}. 

In order to study the contribution from the boundary energy 
functional in the case of boundary at infinity precisely, we 
need to regularize the boundary by taking a large circle with 
the radius $R$ and take the $R\to\infty$ limit. 
The volume $\epsilon^{aib} n_a^\alpha n_b^{\alpha, {\rm bound}}$ 
of the parallerogram formed by the three unit vectors $n_a^\alpha$, 
$n_b^{\alpha, {\rm bound}}$ and the vector $\vec{ e}_i$ tangent 
to the boundary vanishes as $R\to\infty$. 
However, the circumference of the boundary increases as $R\to \infty$. 
As a total derivative, ${E}_{\rm boundary}$ can be 
rewritten into an integral along a circle $C$ with a large radius 
$R$ centered at the location of the skyrmion (defined by $n_3=-1$) 
\begin{eqnarray}
{E}_{\rm boundary}
%&
=%&
-\oint_C dx^i \kappa\; n_i^\alpha 
%\nonumber \\ 
%&
=%&
-\int_0^{2\pi}R d\varphi\; \kappa\; n_\varphi^\alpha 
, 
\label{eq:contour_integral}
\end{eqnarray}
where $n_\varphi^\alpha$ is the tangential component of the 
magnetization vector at the large circle $C$ in the $x_1, x_2$ plane. 
Therefore the boundary energy functional contribute a finite 
result if the magnetization vector decays only as a power of $1/R$ : 
$n_\varphi^\alpha \sim 1/R$. 
It has been noted~\cite{Melcher} that the value of the boundary 
term $\nabla \cdot \left(\vec{n}\times \delta \vec{n}\right)$ 
depends on the fall off conditions on $\vec{n}$ and $\delta\vec{n}$. 
This is discussed from the more general gauged sigma model point 
of view in Ref.~\cite{Schroers5}. In the familiar case of a 
skyrmion energy including just a Zeeman term studied in 
Ref.~\cite{Melcher}, the $A=0$ case of Eq.~\eqref{energy functional static}, 
the fields and their fluctuations fall off exponentially fast 
and the boundary term is zero. However, along the solvable 
line, $B=2A$, the fields and fluctuations fall off more slowly and 
the boundary term is not zero and may not even be well defined 
\cite{Schroers5}. 

To showcase the explicit decay properties of the hedgehog 
configurations consider the radial equation of motion 
\begin{equation}
\frac{d^{2} \Theta}{dr^{2}}=-\frac{1}{r}\frac{d \Theta}{dr}
+\frac{\sin\left(2 \Theta\right)}{2r^{2}}
-2\kappa\sin\left(\alpha+\gamma\right)\frac{\sin^{2} \Theta}{r}
+B\sin  \Theta -A\sin\left(2 \Theta\right). 
\label{general hedgehog profile equation}
\end{equation}
We focus on the case of a magnetic field aligned with the positive 
$z$-axis, $B>0$. 
In particular, we consider the case of $B\geq 2A>0$, where 
the boundary condition is given by $\vec{n}\to 
\vec{n}_{\infty}=(0,0,1)$ 
for large values of $A, B$.
In the limit $r\to \infty$ the boundary condition
on the profile function is $\lim_{r\to \infty} \Theta(r)=0$. 
Considering the limit of \eqref{general hedgehog profile equation} 
and keeping only terms linear in $ \Theta$ we have 
\begin{equation}
\frac{d^{2}  \Theta}{dr^{2}}+\frac{1}{r}\frac{d \Theta}{dr}
-\left(\frac{1}{r^{2}}+\left(B-2A\right)\right) \Theta=0. 
\label{asymptotic profile function equation}
\end{equation}
Asymptotically the hedgehog magnetisation vector becomes 
\begin{equation}
\vec{n}=e_{3}+ \Theta(r)\begin{pmatrix}
\cos\left(\varphi+\gamma\right)\\
 \sin\left(\varphi+\gamma\right)\\
0
\end{pmatrix},
\end{equation}
retaining terms up to first order in $\Theta(r)$ in 
Eq.\eqref{Hedgehog solution}. 
Solving Eq.~\eqref{asymptotic profile function equation} gives
the decay properties of $\vec{n}$.
There are two %three 
cases for this equation related to different forms of Bessel's equation \cite{bowman1958}:
\begin{enumerate}
\item $B>2A$, Equation \eqref{asymptotic profile function equation} is the modified Bessel's equation and thus
\begin{equation}
 \Theta(r)\sim \frac{1}{\sqrt{r}}K_{1}\left(\sqrt{B-2A}r\right)\simeq \frac{1}{\sqrt{r}}e^{-\sqrt{B-2A}r},
\end{equation} 
decaying exponentially fast as $r\to \infty$.
\item $B=2A$, it is straight forward to check that 
$ \Theta(r)\sim \frac{1}{r}$ solves Equation \eqref{asymptotic 
profile function equation}.
\end{enumerate}

This means that for $B=2A$, the magnetisation vector field 
decays slowly and the boundary term is needed to make the variational 
problem well defined. 
For $B>2A$, however, the magnetisation vector field decays 
exponentially fast and the boundary term in the energy 
is zero. 
Therefore the boundary term is mandatory for our solvable line 
$B= 2A$ (the phase boundary between the canted and polarised 
ferromagnetic phases), from the viewpoint of the variational 
principle. 
For $0<B<2A$ (the canted ferromagnetic phase), we have to analyze 
the asymptotic behaviour of the magnetization vector using 
different differential equations from
Eqs.~\eqref{general hedgehog profile equation} and 
\eqref{asymptotic profile function equation}, since the 
boundary condition is now different, $\vec{n}_{\infty}\not=(0,0,1)$.

\section{Expression for the energy density}
\label{interaction energy density computation}

In this appendix, we calculate an expression for the skymion energy in terms of dimensionless parameters. 
The energy density with the boundary term removed is given in 
Eq.~\eqref{energy functional-w}. 
This can be written in terms of two dimensionless parameters 
in the following way. We change coordinates to the dimensionless 
ones $\zeta=\frac{2A}{\kappa}ze^{-i\alpha}$, and 
$\beta=\frac{2A}{\kappa}ze^{-ib}$ in 
Eq.~\eqref{eq:dimensionless-coordinate}. 
Then, the energy density $e[w]$ can be rewritten as

\begin{equation}
e[w]=4\pi \rho_{Q}+\left(\frac{2A}{\kappa}\right)^{2}
\frac{8\vert \partial_{\zeta}\bar{w}\vert^{2}}{\left(1+|w|^{2}\right)^{2}}
+4A\left(\frac{\vert w\vert^{4}
+4\text{Im}\left(w^{2}\partial_{\zeta}\bar{w}\right)}
{\left(1+|w|^{2}\right)^{2}}\right),
\label{energy density rewritten}
\end{equation}

with $\rho_{Q}$ the topological charge density.
The energy is the integral of Eq.~\eqref{energy density rewritten}, 
with the measure scaled as $d^{2}x=\left(\frac{\kappa}{2A}\right)^{2}d^{2}\zeta$ 
under the change of coordinates, it is

\begin{equation}
E[w]=4\pi Q[w]+\mathcal{I}_{1}(|\beta|)+\left(\frac{\kappa^{2}}{2A}\right)
\mathcal{I}_{2}(|\beta|), \label{skyrmion energy expressions}
\end{equation}

where the two dimensionless integrals are defined as 

\begin{align}
\mathcal{I}_{1}(|\beta|)
&=\int \frac{8\vert \partial_{\zeta}\bar{w}\vert^{2}}
{\left(1+|w|^{2}\right)^{2}}d^{2}\zeta, 
\label{dimensionless 1}\\
\mathcal{I}_{2}(|\beta|)&=2\int\left(\frac{\vert w\vert^{4}
+4\text{Im}\left(w^{2}\partial_{\zeta}\bar{w}\right)}
{\left(1+|w|^{2}\right)^{2}}\right)d^{2}\zeta. 
\label{dimensionless 2}
\end{align}

One should note that the functional dependence on the dimensionless 
parameter $\kappa^2/(2A)$ is found to be linear. 
This structure is valid for any field configurations including 
our superposition of two skyrmions. 
The two integrals $\mathcal{I}_{1}, \mathcal{I}_{2}$ for the 
superposition are only functions of the magnitude of the 
dimensionless separation, $|\beta|$.

Applying the formula \eqref{skyrmion energy expressions} 
to the case of a single skyrmion, such as in 
Eq.~\eqref{hedghog sol in complex coords}, 
these integrals can be evaluated explicitly to give
\begin{align*}
\mathcal{I}_{1}(\beta)&=\phantom{-}8\pi,\\
\mathcal{I}_{2}(\beta)&=-8\pi,
\end{align*}
which is consistent with the expression for the energy 
in Eq.~\eqref{hedgehog energy expression}.

Applying the formula \eqref{skyrmion energy expressions} 
to the superposition $w=w_{A}+w_{B}$ and using the 
definition of the interaction energy \eqref{interaction energy}, 
we can write the interaction energy as

\begin{equation}
E_{\text{int}}[w]
=\left[\mathcal{I}_{1}^{\text{sup}}\left(\beta\right)-16\pi\right] 
+\left[\mathcal{I}_{2}^{\text{sup}}\left(\beta\right) +16\pi\right] 
\frac{\kappa^{2}}{2A}
\label{superposition energy appendix}
\end{equation}

where $\mathcal{I}_{1}^{\text{sup}}\left(\beta\right)$ 
and $\mathcal{I}_{2}^{\text{sup}}\left(\beta\right)$ are 
the dimensionless integrals for the superposition field 
$w=w_A+w_B$.

For purely anti-holomorphic functions we can take this a step further and explicitly compute $\mathcal{I}_{1}^{\text{sup}}\left(\beta\right)$. The key observation for this is that in Eq.~\eqref{degree-w} the degree $Q[w]$ is written as
\begin{equation}
Q[w]=\frac{1}{\pi}\int\frac{\vert\partial_{\zeta}w\vert^{2}-\vert\partial_{\zeta}\bar{w}\vert^{2}}{\left(1+\vert w\vert^{2}\right)^{2}}d^{2}\zeta.
\end{equation}
When $w$ is anti-holomorphic, such as for a superposition of skyrmions, it becomes
\begin{equation}
Q[w]=-\frac{1}{\pi}\int\frac{\vert\partial_{\zeta}\bar{w}\vert^{2}}{\left(1+\vert w\vert^{2}\right)^{2}}d^{2}\zeta=-\frac{1}{8\pi}\mathcal{I}_{1}^{\text{sup}}\left(\beta\right).
\end{equation}
The superposition $w=w_{A}+w_{B}$ of two single skyrmions is an anti-holomorphic function with $Q[w]=-2$, as long as $\beta\neq 0$, and thus
\begin{equation}
\mathcal{I}_{1}^{\text{sup}}\left(\beta\right)=-8\pi Q[w]=16\pi.
\end{equation}
This simplifies Eq.~\eqref{superposition energy appendix} to
\begin{equation}
E_{\text{int}}[w]
=\left[\mathcal{I}_{2}^{\text{sup}}\left(\beta\right) +16\pi\right] 
\frac{\kappa^{2}}{2A}. \label{interaction energy appendix}
\end{equation}

%%%%%%%%%%%%%%%
\section{Spiral state}
\label{spiral state appendix}
In this appendix, we discuss spiral states.
The exact spiral solution for $B=A=0$ is
\begin{equation}
\vec{n}=\begin{pmatrix}
0\\
-\sin\left(\kappa x\right)\\
\phantom{-}\cos\left(\kappa x\right)
\end{pmatrix} 
\end{equation}
which has average energy
\begin{equation}
\langle e_{\text{spiral}}\rangle_{B=0} =\frac{1}{L}\int_{-\frac{L}{2}}^{\frac{L}{2}}e_{\text{spiral}} dx=-\frac{\kappa^{2}}{2}.
\end{equation}
The period is defined as $L=\frac{2\pi}{\kappa}$ so that $\langle n_{3}\rangle=0$.
If $A$ is not zero then for Eq.~\eqref{average spiral energy} the boundary conditions imply that

\begin{align}
A\frac{1}{L}\int_{-\frac{L}{2}}^{\frac{L}{2}}\left(1-n_{3}\right)^{2}&=A\frac{1}{L}\int_{-\frac{L}{2}}^{\frac{L}{2}}\left(1-2n_{3}+n_{3}^{2}\right)\nonumber\\
&=A\left(\frac{3}{2}-2\frac{1}{L}\int_{-\frac{L}{2}}^{\frac{L}{2}}\cos(\kappa x)dx+\frac{1}{2}\frac{1}{L}\int_{-\frac{L}{2}}^{\frac{L}{2}}\cos(2\kappa x)dx\right)\nonumber\\
&=\frac{3A}{2}.
\end{align}

This results in
\begin{equation}
\langle e_{\text{spiral}}\rangle =-\frac{\kappa^{2}}{2}+\frac{3A}{2}.
\end{equation}

The energy density of a single skyrmion in a disc of radius $r=c\frac{\kappa}{A}$ is
\begin{equation}
\langle e_{\text{Sk}}\rangle=\frac{4 A}{c^{2}}\left( \frac{A}{\kappa^{2}} -1 \right). \label{average skyrmion energy density general radius}
\end{equation}
Setting this equal to the spiral energy density we find
\begin{equation}
\frac{4 A}{c^{2}}\left( \frac{A}{\kappa^{2}} -1 \right)=-\frac{\kappa^{2}}{2}+\frac{3A}{2},
\end{equation}
which leads to the following quadratic equation for $A$,
\begin{equation}
A^{2}-\frac{8+3c^{2}}{8}\kappa^{2}A+c^{2}\frac{\kappa^{4}}{8}=0,
\end{equation}
solving this for $A$ results in
\begin{equation}
A=\frac{\kappa^{2}}{16}\left(8+3c^{2}\pm\sqrt{64+16c^{2}+9c^{4}}\right).
\end{equation}
As $c$ is positive and we know that $A\leq \kappa^{2}$, as above $A=\kappa^{2}$ we are in the ferromagnetic phase, we take the negative sign and get
\begin{equation}
A=\frac{\kappa^{2}}{16}\left(8+3c^{2}-\sqrt{64+16c^{2}+9c^{4}}\right).
\end{equation}
An interesting choice is to take $c=2$, that is our lattice consists of a single skyrmion in a unit cell which is the disc with radius $r=2\frac{\kappa}{A}$, for which we find
\begin{equation}
A=\frac{\kappa^{2}}{16}\left(20-4\sqrt{17}\right)\simeq 0.22\kappa^{2}
\end{equation}
as the value of $A$ at which we expect a transition between the ferromagnetic phase and the skyrmion lattice. This value of $A$ matches the value that can be read off the phase diagram that was numerically found in \cite{LSB}.


%\begin{thebibliography}{99}
%
%\bibitem{a}
%Author, \emph{Title}, \emph{J. Abbrev.} {\bf vol} (year) pg.
%
%\bibitem{b}
%Author, \emph{Title},
%arxiv:1234.5678.
%
%\bibitem{c}
%Author, \emph{Title},
%Publisher (year).
%
%
%% Please avoid comments such as "For a review'', "For some examples",
%% "and references therein" or move them in the text. In general,
%% please leave only references in the bibliography and move all
%% accessory text in footnotes.
%
%% Also, please have only one work for each \bibitem.
%
%
%\end{thebibliography}
\bibliographystyle{unsrt}
\bibliography{calum-bib}

\begin{thebibliography}{10}

\bibitem{NT}
Naoto Nagaosa and Yoshinori Tokura.
\newblock Topological properties and dynamics of magnetic skyrmions.
\newblock {\em Nature nanotechnology}, 8 12:899--911, 2013.

\bibitem{BY}
A.~Bogdanov and D.~A. Yablonskii.
\newblock Thermodynamically stable vortices in magnetically ordered crystals.
  the mixed state of magnets.
\newblock {\em Zh.~Eksp.~Teor.~Fiz}, 95:178 -- 182, 1989.

\bibitem{YOKPHMNT}
X.~Z. Yu, Yoshinori Onose, Naoya Kanazawa, Joung~Hwan Park,
  Johan~H{\aa}kanssonMengjie Han, Y.~Matsui, Naoto Nagaosa, and Yasuhiro
  Tokura.
\newblock Real-space observation of a two-dimensional skyrmion crystal.
\newblock {\em Nature}, 465:901--904, 2010.

\bibitem{MBJPRNGB}
S.~Muhlbauer, B.~Binz, F.~Jonietz, C.~Pfleiderer, A.~Rosch, A.~Neubauer,
  R.~Georgii, and P.~Boni.
\newblock Skyrmion lattice in a chiral magnet.
\newblock {\em Science}, 323(5916):915–919, Feb 2009.

\bibitem{RHMBWVKW}
Niklas {Romming}, Christian {Hanneken}, Matthias {Menzel}, Jessica~E. {Bickel},
  Boris {Wolter}, Kirsten {von Bergmann}, Andr{\'e} {Kubetzka}, and Roland
  {Wiesendanger}.
\newblock {Writing and Deleting Single Magnetic Skyrmions}.
\newblock {\em Science}, 341(6146):636--639, Aug 2013.

\bibitem{PSZ}
B.~M. A.~G. Piette, B.~J. Schroers, and W.~J. Zakrzewski.
\newblock {Multi - solitons in a two-dimensional Skyrme model}.
\newblock {\em Z. Phys.}, C65:165--174, 1995.

\bibitem{MS}
N.~S. Manton and P.~M. Sutcliffe.
\newblock {\em Topological solitons.}
\newblock Cambridge monographs on mathematical physics. Cambridge University
  Press, Cambridge, July 2004.

\bibitem{Dzyaloshinskii}
I.~Dzyaloshinskii.
\newblock A thermodynamic theory of `weak' ferromagnetism of
  antiferromagnetics.
\newblock {\em J. Phys. Chem. Solids}, 4:241 -- 255, 1958.

\bibitem{Moriya}
T.~Moriya.
\newblock Anisotropic superexchange interaction and weak ferromagnetism,.
\newblock {\em Phys. Rev}, 120:91 -- 98, 1960.

\bibitem{Melcher}
Christof Melcher.
\newblock Chiral skyrmions in the plane.
\newblock {\em Proceedings of The Royal Society A Mathematical Physical and
  Engineering Sciences}, 470, 12 2014.

\bibitem{LSB}
Shi-Zeng {Lin}, Avadh {Saxena}, and Cristian~D. {Batista}.
\newblock {Skyrmion fractionalization and merons in chiral magnets with
  easy-plane anisotropy}.
\newblock {\em Physical Review B}, 91(22):224407, Jun 2015.

\bibitem{BRER}
Sumilan Banerjee, James Rowland, Onur Erten, and Mohit Randeria.
\newblock Enhanced stability of skyrmions in two-dimensional chiral magnets
  with rashba spin-orbit coupling.
\newblock {\em Phys. Rev. X}, 4:031045, Sep 2014.

\bibitem{APL_thinfilms_2016}
Mark Vousden, Maximilian Albert, Marijan Beg, Marc-Antonio Bisotti, Rebecca
  Carey, Dmitri Chernyshenko, David Cortes, Weiwei Wang, Ondrej Hovorka,
  Christopher~H. Marrows, and Hans Fangohr.
\newblock Skyrmions in thin films with easy-plane magnetocrystalline
  anisotropy.
\newblock {\em Applied Physics Letters}, 108(13):1--4, March 2018.

\bibitem{MRG_2016}
Jan Müller, Achim Rosch, and Markus Garst.
\newblock Edge instabilities and skyrmion creation in magnetic layers.
\newblock {\em New Journal of Physics}, 18(6):065006, jun 2016.

\bibitem{BH}
A.~Bogdanov and A.~Hubert.
\newblock Thermodynamically stable magnetic vortex states in magnetic crystals.
\newblock {\em Journal of Magnetism and Magnetic Materials}, 138(3):255 -- 269,
  1994.

\bibitem{RBP}
U.~K. {R{\"o}{\ss}ler}, A.~N. {Bogdanov}, and C.~{Pfleiderer}.
\newblock {Spontaneous skyrmion ground states in magnetic metals}.
\newblock {\em Nature}, 442(7104):797--801, Aug 2006.

\bibitem{HZZJN}
Jung~Hoon Han, Jiadong Zang, Zhihua Yang, Jin-Hong Park, and Naoto Nagaosa.
\newblock Skyrmion lattice in a two-dimensional chiral magnet.
\newblock {\em Phys. Rev. B}, 82:094429, Sep 2010.

\bibitem{KO2015}
Junichiro Kishine and A.S. Ovchinnikov.
\newblock Chapter one - theory of monoaxial chiral helimagnet.
\newblock volume~66 of {\em Solid State Physics}, pages 1 -- 130. Academic
  Press, 2015.

\bibitem{TOPSK_2018}
A.~A. Tereshchenko, A.~S. Ovchinnikov, Igor Proskurin, E.~V. Sinitsyn, and
  Junichiro Kishine.
\newblock Theory of magnetoelastic resonance in a monoaxial chiral helimagnet.
\newblock {\em Physical Review B}, 97(18), May 2018.

\bibitem{BRS}
Bruno Barton-Singer, Calum Ross, and Bernd~J. Schroers.
\newblock {Magnetic Skyrmions at Critical Coupling}.
\newblock {\em Commun. Math. Phys.}, 2020.

\bibitem{BP}
Alexander~M. Polyakov and A.~A. Belavin.
\newblock {Metastable States of Two-Dimensional Isotropic Ferromagnets}.
\newblock {\em JETP Lett.}, 22:245--248, 1975.
\newblock [Pisma Zh. Eksp. Teor. Fiz.22,503(1975)].

\bibitem{MD2016}
Lukas D{\"o}ring and C.~Melcher.
\newblock Compactness results for static and dynamic chiral skyrmions near the
  conformal limit.
\newblock {\em Calculus of Variations and Partial Differential Equations},
  56:60, 2016.

\bibitem{Bogomolny:1975de}
E.~B. Bogomolny.
\newblock {Stability of Classical Solutions}.
\newblock {\em Sov. J. Nucl. Phys.}, 24:449, 1976.
\newblock [Yad. Fiz.24,861(1976)].

\bibitem{Prasad:1975kr}
M.~K. Prasad and Charles~M. Sommerfield.
\newblock {An Exact Classical Solution for the 't Hooft Monopole and the
  Julia-Zee Dyon}.
\newblock {\em Phys. Rev. Lett.}, 35:760--762, 1975.

\bibitem{Eto:2006pg}
Minoru Eto, Youichi Isozumi, Muneto Nitta, Keisuke Ohashi, and Norisuke Sakai.
\newblock {Solitons in the Higgs phase: The Moduli matrix approach}.
\newblock {\em J. Phys.}, A39:R315--R392, 2006.

\bibitem{Shifman:2009zz}
Mikhail Shifman and Alexei Yung.
\newblock {\em {Supersymmetric solitons}}.
\newblock Cambridge Monographs on Mathematical Physics. Cambridge University
  Press, 2009.

\bibitem{Adam:2019yst}
C.~Adam, Jose~M. Queiruga, and A.~Wereszczynski.
\newblock {BPS soliton-impurity models and supersymmetry}.
\newblock {\em JHEP}, 07:164, 2019.

\bibitem{Adam:2019hef}
C.~Adam, K.~Oles, T.~Romanczukiewicz, and A.~Wereszczynski.
\newblock {Domain walls that do not get stuck on impurities}.
\newblock 2019.

\bibitem{Hongo:2019nfr}
Masaru Hongo, Toshiaki Fujimori, Tatsuhiro Misumi, Muneto Nitta, and Norisuke
  Sakai.
\newblock {Instantons in Chiral Magnets}.
\newblock {\em Phys. Rev.}, B101:(in press), 2020.

\bibitem{Schroers4}
Bernd Schroers.
\newblock {Solvable Models for Magnetic Skyrmions}.
\newblock In {\em {11th International Symposium on Quantum Theory and
  Symmetries (QTS2019) Montreal, Canada, July 1-5, 2019}}, 2019.

\bibitem{Schroers5}
Bernd~J Schroers.
\newblock {Gauged Sigma Models and Magnetic Skyrmions}.
\newblock {\em SciPost Phys.}, 7:30, 2019.

\bibitem{Bogdanov1995}
A.~{Bogdanov}.
\newblock {New localized solutions of the nonlinear field equations}.
\newblock {\em JETP Letters}, 62:247, August 1995.

\bibitem{LRSB_2013}
Shi-Zeng Lin, Charles Reichhardt, Cristian~D. Batista, and Avadh Saxena.
\newblock Particle model for skyrmions in metallic chiral magnets: Dynamics,
  pinning, and creep.
\newblock {\em Phys. Rev. B}, 87:214419, Jun 2013.

\bibitem{CGC}
Daniel {Capic}, D.~A. {Garanin}, and E.~M. {Chudnovsky}.
\newblock {Skyrmion-Skyrmion Interaction in a Magnetic Film}.
\newblock Jan 2020.

\bibitem{BLH}
Richard Brearton, Gerrit van~der Laan, and Thorsten Hesjedal.
\newblock Magnetic skyrmion interactions in the micromagnetic framework, 2020.

\bibitem{FKATDS}
David Foster, Charles Kind, Paul~J. Ackerman, Jung-Shen~B. Tai, Mark~R. Dennis,
  and Ivan~I. Smalyukh.
\newblock {Two-dimensional skyrmion bags in liquid crystals and ferromagnets}.
\newblock {\em Nature Phys.}, 15(7):655--659, 2019.

\bibitem{LK2017}
A.~O. Leonov and I.~K\'ezsm\'arki.
\newblock Asymmetric isolated skyrmions in polar magnets with easy-plane
  anisotropy.
\newblock {\em Phys. Rev. B}, 96:014423, Jul 2017.

\bibitem{TSA2019}
V.~E. {Timofeev}, A.~O. {Sorokin}, and D.~N. {Aristov}.
\newblock {Towards an Effective Theory of Skyrmion Crystals}.
\newblock {\em Soviet Journal of Experimental and Theoretical Physics Letters},
  109(3):207--212, February 2019.

\bibitem{SKC}
A.~Saxena, P.~G. Kevrekidis, and J.~Cuevas-Maraver.
\newblock {Nonlinearity and Topology}.
\newblock 2020.

\bibitem{Kobayashi:2014eqa}
Michikazu Kobayashi and Muneto Nitta.
\newblock {Nonrelativistic Nambu-Goldstone modes propagating along a Skyrmion
  line}.
\newblock {\em Phys. Rev.}, D90(2):025010, 2014.

\bibitem{bowman1958}
F.~Bowman.
\newblock {\em Introduction to Bessel Functions}.
\newblock Dover Books on Mathematics. Dover Publications, 1958.

\end{thebibliography}
\end{document}